\documentclass[11pt]{article}

\usepackage[letterpaper, text={6.5in, 9.0in}]{geometry}
\usepackage{amsmath}
\usepackage{amssymb}
\usepackage{amsthm}
\usepackage{color}
\usepackage{graphicx}
\usepackage{bm}
\usepackage{indentfirst}
\usepackage{stmaryrd}
\usepackage{listings}
\theoremstyle{definition}

\newtheorem{theorem}{Theorem}

\newcommand{\RNum}[1]{\uppercase\expandafter{\romannumeral #1\relax}}

\newcommand{\bfzero}[1]{\bm{0}_{#1}}

\newcommand{\calE}{\mathcal{E}}

\newcommand{\bfSigma}{\bm{\Sigma}}
\begin{document}
\title{Extended ADMM for general penalized quantile regression with linear constraints in big data}
\author{Yongxin Liu
\thanks{Corresponding author. School of Statistics and Data Science, Nanjing Audit University, Nanjing, 211815, China (liuyongxin@nau.edu.cn).
}
\and
Peng Zeng\thanks{Department of Mathematics and Statistics, Auburn University, Auburn, AL 36849,
USA (zengpen@auburn.edu).}
       }

\date{}

\maketitle

\begin{abstract}
Quantile regression (QR) can be used to describe the comprehensive relationship between a response and predictors. Prior domain knowledge and assumptions in application are usually formulated as constraints of parameters to improve the estimation efficiency. This paper develops methods based on multi-block ADMM to fit general penalized QR with linear constraints of regression coefficients.
Different formulations to handle the linear constraints and general penalty are explored and compared. The most efficient one has explicit expressions for each parameter and avoids nested-loop iterations in some existing algorithms. Additionally, parallel ADMM algorithm for big data is also developed when data are stored in a distributed fashion. The stopping criterion and convergence of the algorithm are established. Extensive numerical experiments and a real data example demonstrate the computational efficiency of the proposed algorithms. The details of  theoretical proofs and different algorithm variations are presented in Appendix.

\vspace{1em}
\end{abstract}

\section{Introduction}

Quantile regression has gained much attention since the seminal work of \cite{Koenker1978Regression}. As an alternative to mean regression, it models the relationship between the conditional quantile of a response and a set of predictors and provides a more comprehensive picture on the dependence of the response on the predictors.
Given a response vector $y=(y_{1},\ldots,y_{n})^T \in \mathbb{R}^{n}$ and a design matrix $X=(x_{1}, \ldots, x_{n})^T \in \mathbb{R}^{n\times p}$, the coefficients $\beta(\tau)$ in a linear quantile regression can be estimated by
\begin{align*}
\hat\beta(\tau)=\min_{\beta \in \mathbb{R}^{p}}
  \rho_{\tau}(y - X\beta),
\end{align*}
where $\rho_{\tau}(u) = \sum_{i=1}^n \tilde\rho_{\tau}(u_i)$ for $u \in \mathbb{R}^{n}$ and
 $\tilde\rho_{\tau}(z) = \tau z I(z > 0) - (1 - \tau) z I(z \leq 0)$ for $z \in \mathbb{R}$ is the check loss function, where $I(\cdot)$ is the indicator function.
Sometimes, we may write $\hat\beta$ instead of $\hat\beta(\tau)$ for the ease of notation.
A comprehensive review of the quantile regression can be found in \cite{koenker2005quantile}.

In a high-dimensional setting, the predictors are often subject to the sparsity assumption, which is often handled by penalization. Consider a penalized quantile regression problem as follows.
\begin{align*}
\min_{\beta \in \mathbb{R}^{p}} \rho_{\tau}(y - X\beta)+p_{\lambda}(\beta),
\end{align*}
where $p_{\lambda}(\cdot)$ is a penalty function and $\lambda>0$ is a tuning parameter.
Typical penalties include Lasso \cite{Tibshirani1996LASSO}, SCAD \cite{Fan2001SCAD}, MCP \cite{Zhang2010MCP} and so on.
\cite{belloni2011} discussed the theoretical properties of high-dimensional quantile regression with the $\ell_{1}$-penalty.
\cite{wang2012} explored penalized quantile regression in ultra-high dimension for nonconvex penalties such as SCAD and MCP. \cite{li20081} proposed a solution path for the $\ell_1$-norm penalized quantile regression following the LARS algorithm. An iterative coordinate descent algorithm was studied by \cite{Peng2015iterative}.

In many fields, assumptions of parameters or prior knowledge in applications can be formulated in terms of linear equalities or inequalities on $\beta$.
These problems can be found in estimating mechanical structure damage from images \cite{Gorinevsky2009Image},
portfolio selection \cite{Fan2012Vast} and
shape-restricted non-parametric regression \cite{wang2012shape}. In this paper, we consider linearly constrained quantile regression with a general lasso penalty (LCG-QR) as follows.
\numberwithin{equation}{section}
\begin{align}
  \min_{\beta \in \mathbb{R}^{p}} \rho_{\tau}(y - X\beta) +  p_{\lambda}(D\beta), \nonumber\\
  \text{subject to \ $C\beta\geq d$ and $E\beta = f$}, \label{lcgqproblem}
\end{align}
where $\beta \in \mathbb{R}^{p}$ is a vector of unknown parameter to be estimated, $D \in\mathbb{R}^{m\times p}$, $C\in\mathbb{R}^{q\times p}$, $d\in\mathbb{R}^{q}$, $E\in\mathbb{R}^{s\times p}$ and $f\in\mathbb{R}^{s}$ are constant matrices or vectors specified by users according to assumptions or prior knowledge in application. The generalized lasso includes the usual lasso, adaptive lasso \cite{Zou2006Adaptive} and fused lasso \cite{2005Sparsity} as its special cases with proper choices of $D$. \cite{2020Generalized} developed the solution path of $\beta$ for the LCG-QR model, which suits for small or medium-sized data.

In many applications, the data are so large that they can not be loaded into computer memories in a whole. Sometimes, the data are collected and stored in different locations, which may be difficult to move them to one single location. Therefore, there is an urgent need in scalable and distributed methods.
The alternating direction method of multipliers (ADMM) is a popular distributed convex optimization algorithm first introduced by \cite{Gabay1976dual} and \cite{Glowinski1975Sur}.
It can easily be paralleled and implemented in modern distributed computing framework to solve large-scale problems.
A comprehensive discussion on the ADMM can be found in \cite{Boyd2010ADMM}.

Recently, some researchers applied ADMM to solve penalized quantile regression. \cite{yu2017admm}
introduced the ADMM schedule for penalized quantile regression that result in a nested-loop iteration algorithm. \cite{yu2017parallel} proposed a parallel ADMM for large scale data, which is a single-loop algorithm.
\cite{gu2018admm} developed a proximal ADMM algorithm and a sparse coordinate descent ADMM algorithm to solve the quantile regression with folded concave penalties.
\cite{Fan2020Penalized} utilized the slack variable representation to solve penalized quantile regression in big data. However, there is a lack of references to discuss about the algorithms for general penalized quantile regression with linear constraints in model (\ref{lcgqproblem}) for big data.
The constraints of coefficients lead to a stricter solution domain and the general lasso term enforces penalty on the coefficients in a linear combination instead of pure sparsity, which lead to more complex algorithms.

In this paper, we study the methods of fitting LCG-QR model for big data based on the extended ADMM. We explore multiple ways to transfer LCG-QR into the ADMM form and compare their difference. New variables are introduced to represent loss function, general penalty term and linear constraints to avoid inner iteration within the algorithm.
By this way, the ADMM formulation includes three or four blocks of variable, which is different from the aforementioned algorithms for penalized quantile regression.
The direct extension of ADMM to a general case of more than three blocks does not
necessarily converge. Thus, the algorithms based on the extended multi-block ADMM for LCG-QR model should be discussed in detail.

The main contributions of this paper are three-fold. First, we discuss the extended multi-block ADMM and give the sufficient conditions to ensure the convergence.
Second, four ADMM algorithms for LCG-QR model by considering two ways of expressing the linear constraints and general penalty for each are proposed. We compare them from theoretical analysis and numerical experiments.
The most efficient ADMM4.Constr algorithm has explicit expression of each unknown parameter during updating schedule. Third, we investigate the parallel ADMM algorithm for LCG-QR model by splitting the observations into different subsets for big data. This algorithm applies to the scenario when data are too massive to process on a single machine or when data are collected or stored in a distributed fashion.

The remainder of this paper is organized as follows.
In Section 2, we present the extended ADMM for multi-blocks case with an example.
In Section 3, we develop four ADMM algorithms of LCG-QR model and give the convergence results.
In Section 4, we propose the parallel ADMM algorithm for large scale data.
Simulation studies of the ADMM and parallel ADMM algorithm are provided in Section 5. A real data application is made to demonstrate the utility of the method in Section 6.
Finally, we conclude the paper in Section 7.

\section{Extension of ADMM}
In this section, we first review the classic ADMM and then discuss the extension of ADMM to the case of multi-block of variables.

\subsection{Classic Formulation}
The classic ADMM algorithm solves an optimization problem of the following format.
\begin{align}
&\min_{x_{1},x_{2}}  f(x_{1})+g(x_{2}) \nonumber \\
&\text{s.t.}  \quad Ax_{1}+Bx_{2}=c, \label{originalproblem}
\end{align}
where $x_{1} \in \mathbb{R}^{a}$ and $x_{2} \in \mathbb{R}^{b}$ are variables, $A \in \mathbb{R}^{l\times a}$ and $B \in \mathbb{R}^{l\times b}$ are constant matrices, and $c \in \mathbb{R}^{l}$ is a constant vector.
The augmented Lagrangian for (\ref{originalproblem}) is
\begin{align*}
L_{\gamma}(x_{1},x_{2},\eta)=f(x_{1})+g(x_{2})+\eta^{T}(Ax_{1}+Bx_{2}-c)
+\frac{\gamma}{2}\|Ax_{1}+Bx_{2}-c\|_{2}^{2},
\end{align*}
where $\eta \in \mathbb{R}^{l}$ is the dual variable, $\gamma>0$ is an augmented Lagrangian parameter and $\|\cdot\|_{2}$ denotes the $L_{2}$-norm in the Euclidean space. It can be written in a scaled form as
\begin{align*}
L_{\gamma}(x_{1},x_{2},u)=f(x_{1})+g(x_{2})+\frac{\gamma}{2}\|Ax_{1}+Bx_{2}-c+u\|_{2}^{2},
\end{align*}
where $u=\eta/\gamma$ is the scaled dual variable.
The ADMM minimizes $x_{1}$ and $x_{2}$ iteratively,
\begin{align}
x_{1}^{k+1}&=\arg\min_{x_{1}} L_{\gamma}(x_{1},x_{2}^{k},u^{k}),\nonumber \\
x_{2}^{k+1}&=\arg\min_{x_{2}} L_{\gamma}(x_{1}^{k+1},x_{2},u^{k}),\nonumber \\
u^{k+1}&=u^{k}+(Ax_{1}^{k+1}+Bx_{2}^{k+1}-c).  \label{originalADMM}
\end{align}
The advantage of ADMM lies in that the original optimization problem can be solved by a series of optimization problems of much smaller size.

The proposed LCG-QR model with linear constraints and general penalty will need more than three variables to describe the optimization problem, which is different from the classic ADMM.
\cite{Chen2016extension} showed that the direct extension of ADMM to more than three blocks is not necessary convergent and founded sufficient conditions to ensure the convergence.
Therefore, we need to study the extended ADMM in a general multi-block case and explore its convergence result.

\subsection{Direct extension of ADMM}

Consider a convex minimization problem whose objective function is the sum of $N$ functions without coupled variables:
\begin{align}
&\min_{x_{1},\ldots,x_{n}}  \sum_{l=1}^{N}f_{l}(x_{l}) \qquad \text{s.t.}  \ \sum_{l=1}^{N}A_{l}x_{l}=c,  \label{generalproblem}
\end{align}
where $x_{l} \in \mathbb{R}^{a_{l}}$ are variables, $A_{l} \in \mathbb{R}^{h\times a_{l}}$ are constant matrices, $c \in \mathbb{R}^{h}$ is constant vector and $f_{l} : \mathbb{R}^{a_{l}} \rightarrow R$ are closed convex but not necessarily smooth functions.
It is natural to extend (\ref{originalADMM}) directly to obtain the scheme for (\ref{generalproblem})
\begin{align}
\left\{
\begin{aligned}
x_{1}^{k+1}& = \arg\min_{x_{1}} \mathcal{L}_{\gamma}(x_{1},x_{2}^{k},\ldots,x_{N}^{k},u^{k}) ,\\
\vdots \\
x_{N}^{k+1}& = \arg\min_{x_{N}} \mathcal{L}_{\gamma}(x_{1}^{k+1},\ldots,x_{N-1}^{k+1},x_{N},u^{k}) ,\\
u^{k+1}& = u^{k}+(A_{1}x_{1}^{k+1}+\ldots+A_{N}x_{N}^{k+1}-c),
\end{aligned}
\right. \label{extendADMM}
\end{align}
where
\begin{align}
\mathcal{L}_{\gamma}(x_{1},\ldots,x_{N},\eta)=&\sum_{l=1}^{N} f_{l}(x_{l})
+\frac{\gamma}{2}\|A_{1}x_{1}+\ldots+A_{N}x_{N}-c+u\|_{2}^{2}\label{lagfunction}
\end{align}
is the scaled augmented Lagrangian for (\ref{generalproblem}).
If the direct extension of ADMM (\ref{extendADMM}) can be reduced to a special case of the classic ADMM scheme (\ref{originalADMM}) with two blocks of variables, then the
convergence can be guaranteed by existing results in the ADMM literature. Inspired by the approach in \cite{Chen2016extension}, we provide a sufficient condition in the following theorem to ensure the convergence of direct extension of ADMM.

\begin{theorem}
For some constant $M$ less than $N$, the variables $x_{1},x_{2},\ldots,x_{N}$ can be partitioned into two parts $(x_{1},x_{2},\ldots,x_{M})$ and $(x_{M+1},x_{M+2},\ldots,x_{N})$. In each part, if the coefficient matrices of variables are mutually orthogonal, that is,
\begin{align*}
A_{i}^{T}A_{j}=0, \ \text{for}\ i\neq j\in \{1,2,\ldots,M \} \ \text{and}\
A_{k}^{T}A_{l}=0, \ \text{for}\ k\neq l\in \{M+1,M+2,\ldots,N \}.
\end{align*}
Then the direct extension of ADMM in (\ref{extendADMM}) is convergent.
\label{thm:multi-blocks}
\end{theorem}

If there exists some constant $M$ that makes the coefficient matrices satisfy the condition in Theorem 1, then $(x_{1},x_{2},\ldots,x_{M})$ can be regarded as one variable and $(x_{M+1},x_{M+2},\ldots,x_{N})$ can be regarded as one variable in the classic ADMM, thus the direct extension of ADMM (\ref{extendADMM}) is convergent. The proof of Theorem 1 is given in Appendix A.

\subsection{An example of extended ADMM}

We take a convex minimization model where the objective function consists of four functions as an example to show the extended ADMM,
\begin{align}
&\min_{x_{1},x_{2},x_{3},x_{4}} f_{1}(x_{1})+f_{2}(x_{2})+f_{3}(x_{3})+f_{4}(x_{4}) \nonumber \\
&\text{s.t.}  \quad A_{1}x_{1}+A_{2}x_{2}+A_{3}x_{3}+A_{4}x_{4}=c.  \label{exampleproblem}
\end{align}
The updates of variables $(x_{1},x_{2},x_{3},x_{4})$ follows the direct extension form in (\ref{extendADMM}). We present the cases where the algorithm is convergent according to Theorem 1:
\begin{itemize}
\item
Case 1: $A_{2}^{T}A_{3}=0$, $A_{2}^{T}A_{4}=0$, $A_{3}^{T}A_{4}=0$;
\item
Case 2: $A_{1}^{T}A_{2}=0$, $A_{3}^{T}A_{4}=0$;
\item
Case 3: $A_{1}^{T}A_{2}=0$, $A_{1}^{T}A_{3}=0$, $A_{2}^{T}A_{3}=0$.
\end{itemize}
If one of the above three cases holds, the extended ADMM is convergent.
According to \cite{Boyd2010ADMM}, we illustrate the stopping criterion of the algorithm.
The primal residual is
\begin{align*}
r_{\text{pri}}^{k+1}=A_{1}x_{1}^{k+1}+A_{2}x_{2}^{k+1}+A_{3}x_{3}^{k+1}+A_{4}x_{4}^{k+1}-c,
\end{align*}
and the dual residuals are
\begin{align*}
s_{1}^{k+1}&=\gamma A_{3}^{T}A_{4}(x_{4}^{k+1}-x_{4}^{k}),\\
s_{2}^{k+1}&=\gamma A_{2}^{T}A_{3}(x_{3}^{k+1}-x_{3}^{k})+\gamma A_{2}^{T}A_{4}(x_{4}^{k+1}-x_{4}^{k}),\\
s_{3}^{k+1}&=\gamma A_{1}^{T}A_{2}(x_{2}^{k+1}-x_{2}^{k})+\gamma A_{1}^{T}A_{3}(x_{3}^{k+1}-x_{3}^{k})
+\gamma A_{1}^{T}A_{4}(x_{4}^{k+1}-x_{4}^{k}).
\end{align*}
A reasonable stopping criterion is that the primal and dual residuals are small, i.e.,
\begin{align*}
\|r_{\text{pri}}^{k+1}\|_{2}\leq \epsilon^{\text{pri}}, \ \|s_{1}^{k+1}\|_{2}\leq \epsilon_{1}^{\text{dual}}, \ \|s_{2}^{k+1}\|_{2}\leq \epsilon_{2}^{\text{dual}}, \ \|s_{3}^{k+1}\|_{2}\leq \epsilon_{3}^{\text{dual}},
\end{align*}
where $\epsilon^{\text{pri}}>0$ and $\epsilon_{1}^{\text{dual}}>0$, $\epsilon_{2}^{\text{dual}}>0$, $\epsilon_{3}^{\text{dual}}>0$ are feasibility tolerances
and can be chosen using an absolute and relative criterion, such as
\begin{align*}
\epsilon ^{\text{pri}} &=\sqrt{l}\epsilon ^{\text{abs}}+\epsilon ^{\text{rel}}\max\{\|A_{1}x_{1}^{k+1}\|_{2}, \|A_{2}x_{2}^{k+1}\|_{2}, \|A_{3}x_{3}^{k+1}\|_{2},
\|A_{4}x_{4}^{k+1}\|_{2}, \|c\|_{2}\},\\
\epsilon_{1} ^{\text{dual}} &=\sqrt{a_{3}}\epsilon ^{\text{abs}}+\gamma\epsilon ^{\text{rel}}\|A_{3}^{T}u^{k+1}\|_{2}, \\
\epsilon_{2} ^{\text{dual}} &=\sqrt{a_{2}}\epsilon ^{\text{abs}}+\gamma\epsilon ^{\text{rel}}\|A_{2}^{T}u^{k+1}\|_{2}, \\
\epsilon_{3} ^{\text{dual}} &=\sqrt{a_{1}}\epsilon ^{\text{abs}}+\gamma\epsilon ^{\text{rel}}\|A_{1}^{T}u^{k+1}\|_{2}.
\end{align*}
The choice of $\epsilon ^{\text{abs}}$ and $\epsilon ^{\text{rel}}$ depend on the application, which
might be $\epsilon ^{\text{abs}}=\epsilon ^{\text{rel}} = 10^{-2}$ or $10^{-3}$.
The derivation of the residuals and stopping criterion is given in Appendix A.

\section{LCG-QR ADMM Algorithms}
The main difference between model (\ref{lcgqproblem}) and traditional penalized quantile regression is that it includes linear constraints and general penalty. Therefore, we focus on these two issues through multiple ways of formulating the problem into ADMM form, called LCG-QR ADMM algorithms.
The first one is how to treat the linear constraints. One possible way is to add an indicator function in the objective function, which is 0 if the constraints holds or infinity if not. This approach forces the algorithm only searches the feasible region where the constraints hold.
An alterative way is to incorporate the linear constraints directly in the ADMM formulation. Although the algorithm may search minimizer beyond the feasible region in the beginning, the linear constraints will be satisfied when the algorithm converges.
The second issue to consider is how to handle the penalty part.
We may consider it together with the loss function, which leads to three blocks of variables in ADMM. Alternatively, we may introduce one more variable for the penalty part, which leads to four blocks of variables.

As a summary, two issues with two solutions for each lead to four possible ways of implementation, which are denoted by ADMM4.Constr, ADMM4.Proj, ADMM3.Constr and ADMM3.Proj, respectively. The computational efficiency of four algorithms is not the same and we will compare them carefully. In this section we illustrate the most efficient ADMM4.Constr algorithm in detail and the remaining three algorithms are postponed in Appendix B.

We now introduce new variables $r \in \mathbb{R}^{n}$, $z \in \mathbb{R}^{m}$, $w \in \mathbb{R}^{q}$
and write (\ref{lcgqproblem}) as
  \begin{align}
  &\min_{\beta,r,z,w} \
  \rho_{\tau}(r) + p_{\lambda}(z)+\phi(w),\nonumber \\
  &\text{subject to \ $y-X\beta=r$, $D\beta=z$, $C\beta-w=d$, $E\beta = f$},\label{newproblem}
\end{align}
where $\phi(w)=0$ if $w\geq 0$ componentwise and $=\infty$ otherwise. Thus, there are four unknown variables needed to be estimated.
Denote \begin{align*}
A_{1}=
 \begin{pmatrix}
 X \\
 D \\
 C \\
  E
 \end{pmatrix}, \quad
A_{2}
 = \begin{pmatrix}
  I_{n} \\
  0_{m} \\
  0_{q} \\
  0_{s}
 \end{pmatrix},\quad
 A_{3}
 = \begin{pmatrix}
  0_{n} \\
  -I_{m} \\
  0_{q} \\
  0_{s}
 \end{pmatrix},\quad
 A_{4}
 = \begin{pmatrix}
  0_{n} \\
  0_{m} \\
  -I_{q} \\
  0_{s}
 \end{pmatrix},\quad
c=\begin{pmatrix}
  y\\
  0_{m} \\
  d \\
  f
 \end{pmatrix}.
\end{align*}
It has the relationship $A_{1}\beta+A_{2}r+A_{3}z+A_{4}w=c$.
The augmented Lagrangian for (\ref{newproblem}) is
\begin{align*}
\mathcal{L}(\beta,r,z,w,u)
= \rho_{\tau}(r) +  p_{\lambda}(z)+\phi(w)
  +\frac{\gamma}{2}\|A_{1}\beta+A_{2}r+A_{3}z+A_{4}w-c+u\|_{2}^{2},
\end{align*}
where $u=(u_{1}^{T},u_{2}^{T},u_{3}^{T},u_{4}^{T})^{T}$ is the scaled dual variable satisfying $u_{1} \in \mathbb{R}^{n}$,
$u_{2} \in \mathbb{R}^{m}$, $u_{3} \in \mathbb{R}^{q}$, $u_{4} \in \mathbb{R}^{s}$.
Following (\ref{extendADMM}), the ADMM updating rules is as follows,
\begin{align}
\beta^{k+1}&=\arg\min_{\beta}L(\beta,r^{k},z^{k},w^{k},u^{k}),\\
r^{k+1}&=\arg\min_{r}L(\beta^{k+1},r,z^{k},w^{k},u^{k}),\\
z^{k+1}&=\arg\min_{z}L(\beta^{k+1},r^{k+1},z,w^{k},u^{k}),\\
w^{k+1}&=\arg\min_{w}L(\beta^{k+1},r^{k+1},z^{k+1},w,u^{k}),\\
u^{k+1}&=u^{k}+(A_{1}\beta^{k+1}+A_{2}r^{k+1}+A_{3}z^{k+1}+A_{4}w^{k+1}-c).
\end{align}
In (3.2), the problem is equivalent to the minimization of the function
\begin{align*}
\arg\min_{\beta}\|A_{1}\beta+A_{2}r^{(k)}+A_{3}z^{(k)}+A_{4}w^{(k)}-c+u^{(k)}\|_{2}^{2},
\end{align*}
which is a least squares problem of $\beta$ and has the solution
\begin{align*}
\beta^{(k+1)}
=(X^{T}X+D^{T}D+C^{T}C+E^{T}E)^{-1}&\big(X^{T}(y-r^{(k)}-u_{1}^{(k)})+
D^{T}(z^{(k)}-u_{2}^{(k)})\\
&+C^{T}( d+w^{(k)}-u_{3}^{(k)})+E^{T}(f-u_{4}^{(k)})\big).
 \end{align*}
 In (3.3), after discarding the terms independent of $r$, we need to minimize
\begin{align*}
 \rho_{\tau}(r)+\frac{\gamma}{2}
\|X\beta^{(k+1)}+r-y+u_{1}^{(k)}\|_{2}^{2}.
\end{align*}
It can be solved componentwise and has the following soft-thresholding solution,
\begin{align*}
r_{i}^{(k+1)}
=[y_{i}-x_{i}^{T}\beta^{(k+1)}-u_{1i}^{(k)}-\tau/\gamma]_{+}-
[-y_{i}+x_{i}^{T}\beta^{(k+1)}+u_{1i}^{(k)}+(\tau-1)/\gamma]_{+}.
\end{align*}
Here, $(x)_{+}=x$ if $x>0$ and =0, otherwise.
In (3.4), the optimization is equivalent to minimize
\begin{align*}
  p_{\lambda}(z)+\frac{\gamma}{2}\|D\beta^{(k+1)}-z+u_{2}^{(k)}\|_{2}^{2}.
\end{align*}
Assume the penalty term has the additive form $p_{\lambda}(z)=\sum\limits_{j=1}^{m}p_{\lambda}(z_{j})$, which holds for most common penalties, including Lasso, SCAD and MCP. Then the above problem can be solved componentwise and for the Lasso penalty the solution is
\begin{align*}
z_{j}^{(k+1)}
 =(D_{j}\beta^{(k+1)}+u_{2j}^{(k)}-\lambda/\gamma)_{+}
 -(-D_{j}\beta^{(k+1)}-u_{2j}^{(k)}-\lambda/\gamma)_{+}.
 \end{align*}
The solution for other penalties such as SCAD and MCP are postponed in Appendix C.\\
In (3.6), after discarding the terms independent of $w$, we need to solve the problem
\begin{align*}
\min_{w}\ \phi(w)+\frac{\gamma}{2}\|C\beta^{(k+1)}-w-d +u_{3}^{(k)}\|_{2}^{2}.
\end{align*}
The solution of this problem is $( C\beta^{(k+1)}-d +u_{3}^{(k)})_{+}$.

Finally, the update of $u$ is given in (3.6).

We define the primal and dual residuals of the algorithm as follows due to the fact $A_{2}^{T}A_{3}=0$, $A_{2}^{T}A_{4}=0$, $A_{3}^{T}A_{4}=0$,
  \begin{align*}
  r_{\text{pri}}^{k+1}
  =\begin{pmatrix}
 X\beta^{(k+1)}+r^{(k+1)}-y \\
 D\beta^{(k+1)}-z^{(k+1)} \\
 C\beta^{(k+1)}-w^{(k+1)}-d \\
 E\beta^{(k+1)}-f
 \end{pmatrix}
   \end{align*}
and
 \begin{align*}
s^{k+1}=\gamma\big(X^{T}(r^{(k+1)}-r^{(k)})-D^{T}( z^{(k+1)}-z^{(k)})-C^{T}( w^{(k+1)}-w^{(k)})\big).
\end{align*}
The algorithm stops when $r_{\text{pri}}^{k+1}$ and $s^{k+1}$ are small enough, i.e.,
\begin{align*}
\|r_{\text{pri}}^{k+1}\|_{2}\leq \epsilon^{\text{pri}}, \quad \|s^{k+1}\|_{2}\leq \epsilon^{\text{dual}}.
\end{align*}
 These tolerances $\epsilon^{\text{pri}}$ and $\epsilon^{\text{dual}}$ can be chosen using an absolute and relative criterion, such as
 \begin{align*}
 \epsilon ^{\text{pri}} &=\sqrt{n+m+q+s}\epsilon ^{\text{abs}}+\epsilon ^{\text{rel}}\max\{\|A_{1}\beta^{(k+1)}\|_{2}, \|A_{2}r^{(k+1)}\|_{2},  \|A_{3}z^{(k+1)}\|_{2},
\|A_{4}w^{(k+1)}\|_{2}, \|c\|_{2}\},\\
 \epsilon^{\text{dual}} &=\sqrt{p}\epsilon ^{\text{abs}}+\gamma\epsilon ^{\text{rel}}\|A_{1}^{T}u^{k+1}\|_{2}.
 \end{align*}

 Based on the above discussion, the ADMM4.Constr algorithm for problem (\ref{newproblem}) consists of the following steps:

Step 1. Find initial estimate $\beta^{0}$ from least squares regression using $y$ and $X$. Let the initial estimates
$r^{0}=y-X\beta^{0}$, $z^{0}=D\beta^{0}$, $w^{0}=C\beta^{0}-d$ and $u^{0}=0$.

Step 2. At iteration $k+1$, update $(\beta^{(k+1)},r^{(k+1)},z^{(k+1)},w^{(k+1)}, u^{(k+1)})$ by the expressions described above.

Step 3. Stop the algorithm if the stopping criterion is met at step $k+1$. Then $(\beta^{(k+1)},r^{(k+1)},$\\
$z^{(k+1)},w^{(k+1)}, u^{(k+1)})$ are final estimates. Otherwise, go to step 2.

Note that in the existing ADMM algorithms for penalized quantile regression, e.g. \cite{yu2017admm}, \cite{yu2017parallel} and \cite{gu2018admm}, the update for $\beta$ is a lasso type problem. It should be solved by the iterative method such as coordinate descent (CD), which results in a nested-loop scheme. Compared to these algorithms, our method avoids inner iteration by introducing new variable $z=D\beta$.
The $\beta$-update is a least square problem and have a closed-form formula. In fact, all the unknown parameters have explicit expressions in the proposed ADMM4.Constr, which reduces the computational complexity and is much easier to implement.

The ADMM4.Proj algorithm also includes four blocks of variables but the update of $w$ needs to find the projection onto a polyhedra, which has no closed form in general. From this point of view, ADMM4.Constr is better than
ADMM4.Proj. Both the ADMM3.Constr and ADMM3.Proj algorithms include three blocks of variables
and the update for $\beta$ is a generalized lasso problem. It requires numerical optimization, which results in a nested-loop algorithm. Therefore, the ADMM4 algorithms are more efficient than the ADMM3 algorithms.

The coefficient matrices of variables in the above four ADMM algorithms satisfy the condition for direct extension of ADMM as we discussed in Section 2.2 and 2.3.
The convergence of ADMM for convex problems has been well established in the literature.
It is easy to check the assumptions in \cite{Boyd2010ADMM} and \cite{Mota2011Convergence} hold for (\ref{newproblem}) with convex penalties.
Thus the convergence of LCG-QR ADMM algorithm is guaranteed and summarized in the following theorem.

\begin{theorem}
For convex penalties $p_{\lambda}$, the LCG-QR ADMM algorithm converges to the solution of problem (\ref{newproblem}), that is, $\beta^{k}$ generated by the iterative LCG-QR ADMM algorithm converges to a point $\beta^{*}$ that solves (\ref{newproblem}).
\end{theorem}
The convergence of ADMM for nonconvex penalties is a stick problem due to the lack of convexity. Although some scholars have discussed this issue, it has not been solved. The theoretical convergence of the proposed ADMM with nonconvex penalties still needs further study.

\section{Parallelization of LCG-QR ADMM }

When data are too large for a single computer to score or process, distributed computing on multi-computers is a good solution. The ADMM is suitable for computing large-scale optimization problems by decomposing the original problem into a series of sub-problems that can be calculated in parallel.
In this section, we first briefly describe how a distributed algorithm works in general, and then show the parallelization of the most efficient ADMM4.Constr algorithm with Lasso penalty. The other three algorithms can be derived similarly.

A distributed algorithm is designed to be implemented and executed in a system consisting of multiple relatively independent computing units, which are referred to as local machines in this article.
Such a system can be a single computer with multiple cores, a cluster with several computers, or a supercomputer with many nodes.
The most important feature of a local machine is that it has own CPU and memory, which makes it possible to process or compute its own share of data independently from other local machines.
There exists a protocol among the local machines which allows the local machines to communicate with each other.
The communication means that some data can be exchanged among the local machines.
It is important to point out that the data exchanged among the local machines are not the original raw data, but the intermediate numbers calculated during the computing process.
Roughly speaking, in a typical distributed algorithm, each local machine independently processes its own data, intermediate results are collected and aggregated together to produce a global and final result. If necessary, the global result can be sent back to each local machine as an initial value to start another iteration.

Assume that the whole data are stored in $M$ local machines and the $m$th machine contains a subset of $n_{m}$ observations denoted as $\{y_{m},X_{m}\}$, $m=1,\ldots,M$. Denote
$y=(y_{1},\ldots,y_{M})$, $X=(X_{1},\ldots,X_{M})$ and $\sum_{m=1}^{M}n_{m}=n$.
The objective function of penalized quantile regression is $\sum\limits_{m=1}^{M}
\rho_{\tau}(y_{m}-X_{m}\beta) + \lambda\|D\beta\|_{1}$. To solve this global problem in a parallel way, we let $\beta_{m}$ denote the local parameter for the $m$th subset and impose a global constraint $\beta_{m}=\beta$. In addition, we introduce the local variable $r_{m}$ and global variable $z$, $w$, then the LCG-QR problem for distributed data can be written as
\begin{align}
  &\min_{\beta_{m},r_{m},z,w,\beta} \sum_{m=1}^{M}
  \rho_{\tau}(r_{m}) + \lambda\|z\|_{1}+g(w),\nonumber \\
  &\text{subject to} \quad y_{m}-X_{m}\beta_{m}=r_{m}, \ \beta_{m}=\beta,\nonumber \\
  &D\beta_{m}=z,\ C\beta_{m}-w=d, \ E\beta_{m} = f, \quad m=1,2,\ldots,M. \label{parallelproblem}
\end{align}
We solve problem (\ref{parallelproblem}) by the parallel ADMM with the following updates.
\begin{align*}
\beta_{m}^{(k+1)}
&=\arg\min_{\beta_{m}}\|
 X_{m}\beta_{m}+r_{m}^{(k)}- y_{m}+u_{m1}^{(k)}\|_{2}^{2}
 +\|D\beta_{m}-z^{(k)}+u_{m2}^{(k)}\|_{2}^{2}\\
& +\|C\beta_{m}-w^{(k)}-d+u_{m3}^{(k)}\|_{2}^{2}
 +\|E\beta_{m}-f+u_{m4}^{(k)}\|_{2}^{2}
 +\|\beta_{m}- \beta^{(k)}+u_{m5}^{(k)}\|_{2}^{2},\\
r_{m}^{(k+1)}
&= \arg\min_{r_{m}}  \rho_{\tau}(r_{m})+\frac{\gamma}{2}
\|X_{m}\beta_{m}^{(k+1)}+r_{m}-y_{m}+u_{m1}^{(k)}\|_{2}^{2},\\
z^{(k+1)}
&=\arg\min_{z} \lambda\|z\|_{1}+\frac{\gamma}{2}\sum_{m=1}^{M}\|D\beta_{m}^{(k+1)}-z+u_{m2}^{(k)}\|_{2}^{2},\\
w^{(k+1)}&=\arg\min_{w} g(w)+\frac{\gamma}{2}\sum_{m=1}^{M}\|
 C\beta_{m}^{(k+1)}-w-d +u_{m3}^{(k)}\|_{2}^{2},\\
\beta^{(k+1)}&= \arg\min_{\beta} \sum_{m=1}^{M} \|\beta_{m}^{(k+1)}-\beta+u_{m5}^{(k)}\|_{2}^{2}.
\end{align*}
The dual variable vector $u$ is updated as follows,
\begin{align*}
&u_{m1}^{(k+1)}
=u_{m1}^{(k)}+X_{m}\beta_{m}^{(k+1)}+r_{m}^{(k+1)}-y_{m}, \\
&u_{m2}^{(k+1)}
=u_{m2}^{(k)}+D\beta_{m}^{(k+1)}-z^{(k+1)}, \\
&u_{m3}^{(k+1)}
=u_{m3}^{(k)}+ C\beta_{m}^{(k+1)}-w^{(k+1)}-d, \\
&u_{m4}^{(k+1)}
=u_{m4}^{(k)}+ E\beta_{m}^{(k+1)}-f, \\
&u_{m5}^{(k+1)}
=u_{m5}^{(k)}+\beta_{m}^{(k+1)}-\beta^{(k+1)}.
\end{align*}
The update for the local parameter $\beta_{m}$ and $r_{m}$ are based on the observations only on the $m$-th machine. Thus they can be computed on each machine separately. The update for $\beta_{m}$ is a least squares problem and the solution is
\begin{align*}
\beta_{m}^{(k+1)}
=(X_{m}^{T}X_{m}+D^{T}D+C^{T}C+E^{T}&E+I_{p})^{-1}\cdot
 \big(X_{m}^{T}(y_{m}-r_{m}^{(k)}-u_{m1}^{(k)})+D^{T}(z^{(k)}-u_{m2}^{(k)})\\
 &+C^{T}(w^{(k)}+d-u_{m3}^{(k)})+E^{T}(f-u_{m4}^{(k)})+ \beta^{(k)}-u_{m5}^{(k)}\big).
\end{align*}
The update for $r_{m}$ can be solved componentwise and for the $i$th component it is
\begin{align*}
{r_{m}}_{i}^{(k+1)}
=\big[{y_{m}}_{i}-{x_{m}}_{i}^{T}\beta_{m}^{(k+1)}-{u_{m1}}_{i}^{(k)}-\gamma^{-1}\tau\big]_{+}-
\big[-{y_{m}}_{i}+{x_{m}}_{i}^{T}\beta_{m}^{(k+1)}+{u_{m1}}_{i}^{(k)}+\gamma^{-1}(\tau-1)\big]_{+}.
\end{align*}
For the global variables $z$, $w$ and $\beta$, the update should be aggregated by the local parameters $\beta_{m}$ and $u_{m}$ in all machines. The update for $z$ can be solved componentwise and has the form
\begin{align*}
z_{j}^{(k+1)}
 =\big[\frac{1}{M}\sum_{m=1}^{M}(D_{j}\beta_{m}^{(k+1)}+{u_{m2}}_{j}^{(k)})-\lambda/\gamma M\big]_{+}
 -\big[-\frac{1}{M}\sum_{m=1}^{M}(D_{j}\beta_{m}^{(k+1)}+{u_{m2}}_{j}^{(k)})-\lambda/\gamma M\big]_{+}.
 \end{align*}
The update for $w$ is
\begin{align*}
w^{(k+1)}
 = \big[\frac{1}{M} \sum_{m=1}^{M}(C\beta_{m}^{(k+1)}-d +u_{m3}^{(k)})\big]_{+}.
  \end{align*}
The global regression parameter $\beta$ is conducted by a center machine through all $(\beta_{m},u_{m})$ values, and the centre broadcasts the updated $\beta$ back to each machine updating.
\begin{align*}
\beta^{(k+1)}=\frac{1}{M}\sum_{m=1}^{M}(\beta_{m}^{(k)}+u_{m5}^{(k)}).
\end{align*}
Similar as the ADMM4.Constr algorithm, the update of parallel ADMM is stopped when the primal and dual residuals are small enough, the details are postponed in Appendix C.

\section{Simulation}

In this section, we use several simulation examples to illustrate the performance of the above four LCG-QR ADMM algorithms and the parallel algorithm. All the experiments are accomplished in R on a Windows computer with the Intel Core i5 2.40 GHz CPU. All the simulation results are based on 100 independent replications.

\subsection{Simulation for LCG-QR ADMM}
We first compare the performance of the above four LCG-QR ADMM algorithms and then investigate the finite-sample performance for the most efficient algorithm. The model setup is similar to that in \cite{Peng2015iterative}. First generate independent variables $(\widetilde{X}_{1}, \widetilde{X}_{2}, \ldots, \widetilde{X}_{p})$ from $\mathcal{N}(\bfzero{p},\bfSigma)$, where $\bfSigma$ is the covariance matrix with elements $\sigma_{ij}=0.5^{|i-j|}$, $1 \leq i, j \leq p$. Then set $X_{1}=\Phi(\widetilde{X}_{1})$ and $X_{k}=\widetilde{X}_{k}$ for $k=2, 3, \ldots, p$, where $\Phi$ is the cumulative distribution function of the standard normal distribution. Consider the following heteroscedastic regression model,
\begin{align}
 Y=X_{5}+X_{6}+X_{11}+X_{12}+X_{1}\epsilon, \label{experiment1}
\end{align}
where the random error $\epsilon \sim \mathcal{N}(0,1)$.
The quantile regression coefficient of model (\ref{experiment1}) is $\beta=(\epsilon_{\tau},0,0,0,1,1,0,0,0,0,$ $1,1,0,\ldots,0)$, where $\epsilon_{\tau}$ is the $\tau$th sample quantile of $\epsilon$.
The adjacent variables $X_{j}$ in model (\ref{experiment1}) have the same effect on $Y$, but the variable $X_{1}$ affects the quantiles of the response $Y$ except for the 0.5-th quantile.
Generate independent samples $\{(x_{i}, y_{i}), i=1,\ldots,n\}$ from (\ref{experiment1}) of sample size $n$, where $x_{i}$ is the vector of independent variables. We estimate the parameter $\beta=(\beta_{1},\beta_{2},\ldots,\beta_{p})$ for a given $\lambda$ as,
\begin{align}
\min_{\beta \in R^{p}} &\sum_{i=1}^{n}\tilde\rho_{\tau}(y_{i}-x_{i}^{T}\beta)+\lambda\sum_{j=1}^{p}|\beta_{j}|+\lambda\sum_{j=2}^{p}|\beta_{j}-\beta_{j-1}|\nonumber  \\
 &\text{s.t.} \ \beta_{5} \geq 0,  \ \beta_{6} \geq 0, \ \beta_{11} \geq 0,
   \ \beta_{12} \geq 0, \nonumber  \\
   & \quad -3\beta_{5}+\beta_{10}+\beta_{12}+\beta_{15}=-1.\label{experiment2}
\end{align}

\subsubsection{Comparison of four ADMM algorithms}
The aforementioned four ADMM algorithms solve the same optimization problem (\ref{lcgqproblem}), which should lead to the same solution. Therefore, we compare the computational efficiency of these algorithms through the convergence speed and average computing time.

We first explore the convergence performance of algorithms with the increase of iteration step.
The objective function values $\sum\limits_{i=1}^{n} \tilde\rho_{\tau}(y_{i} - x_{i}^{T}\widehat{\beta}(\lambda))/n$ fitted by ADMM4.Constr, ADMM4.Proj, ADMM3.Constr and ADMM3.Proj for $\tau=0.5$, $\lambda=0.5$ are shown on the left of Figures 1-3. Notice that model (\ref{experiment1}) should satisfy the equality and inequality constraints of $\beta$. The simulation results show that the inequality constraints are easy to met, thus we only record
the equality constraint values $\|E\widehat{\beta}(\lambda)-f\|_{1}$ on the right of Figures 1-3.
It can be seen that the objective function of ADMM4.Constr and ADMM3.Constr converge faster than ADMM4.Proj and ADMM3.Proj and they yield almost the same value after enough iteration steps. Meanwhile, the ADMM4.Constr and ADMM3.Constr reach equality constraint faster than ADMM4.Proj and ADMM3.Proj.

 \begin{figure}[h!]
 \centering
\includegraphics[height=0.48\textwidth]{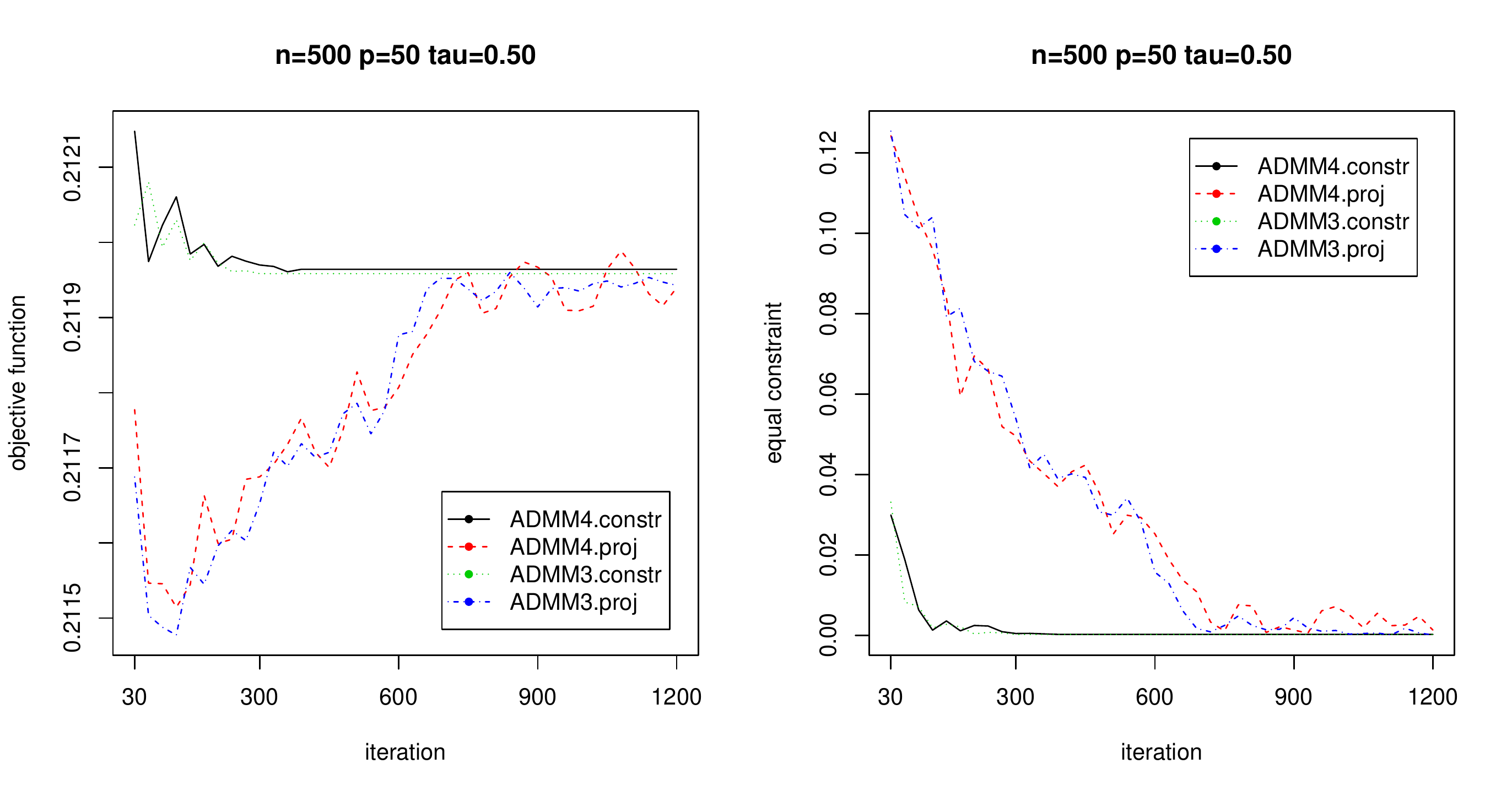}
\small
  \caption{ Objective function values and equal constraint values of four LCG-QR ADMM algorithms versus iteration steps for $n=500 \ p=50$ and $\tau=0.50$.  }
\end{figure}

 \begin{figure}[h!]
 \centering
\includegraphics[height=0.5\textwidth]{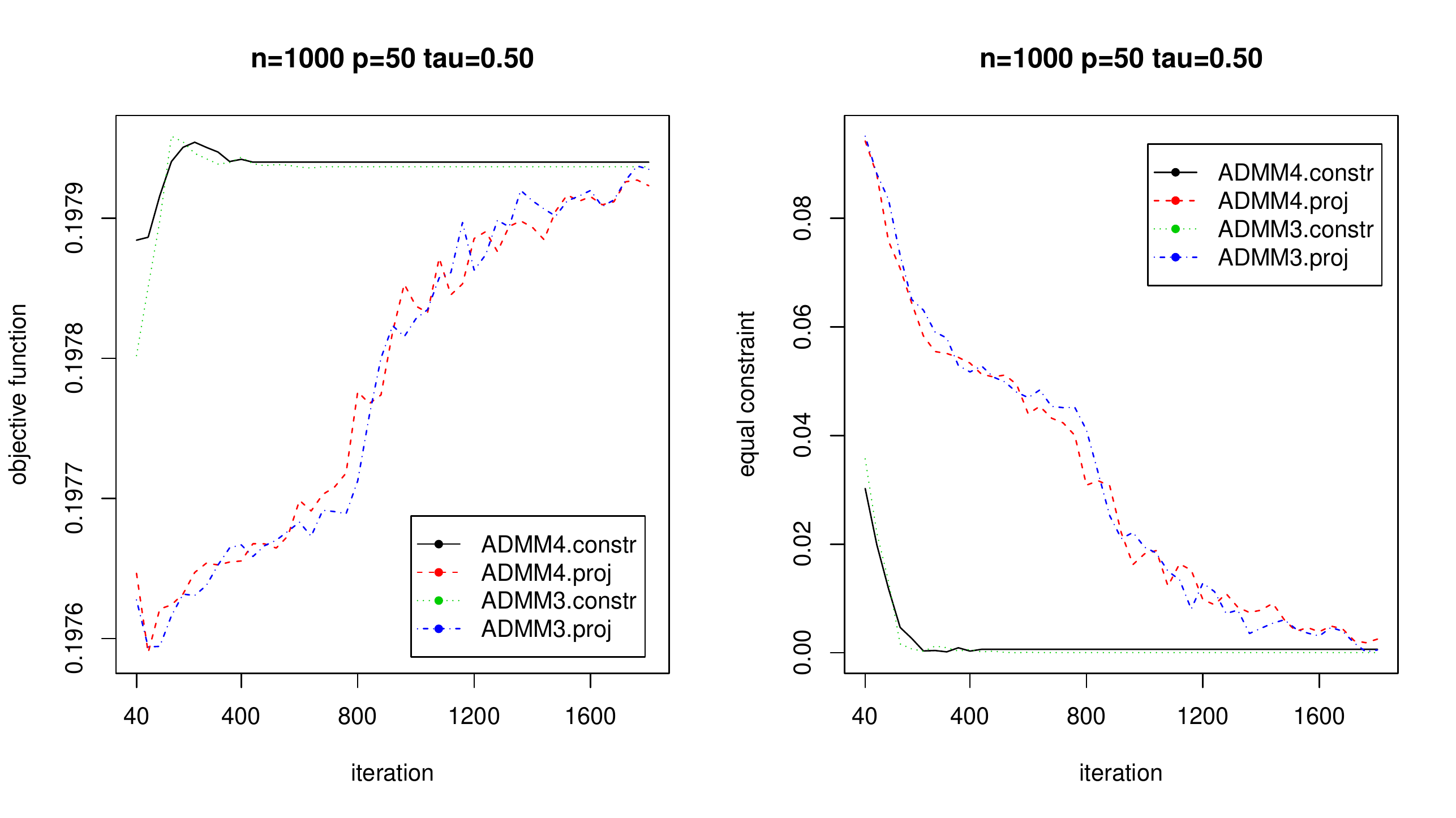}
\small
  \caption{ Objective function values and equal constraint values of four LCG-QR ADMM algorithms versus iteration steps for $n=1000 \ p=50$ and $\tau=0.50$.  }
\end{figure}

 \begin{figure}[h!]
 \centering
\includegraphics[height=0.47\textwidth]{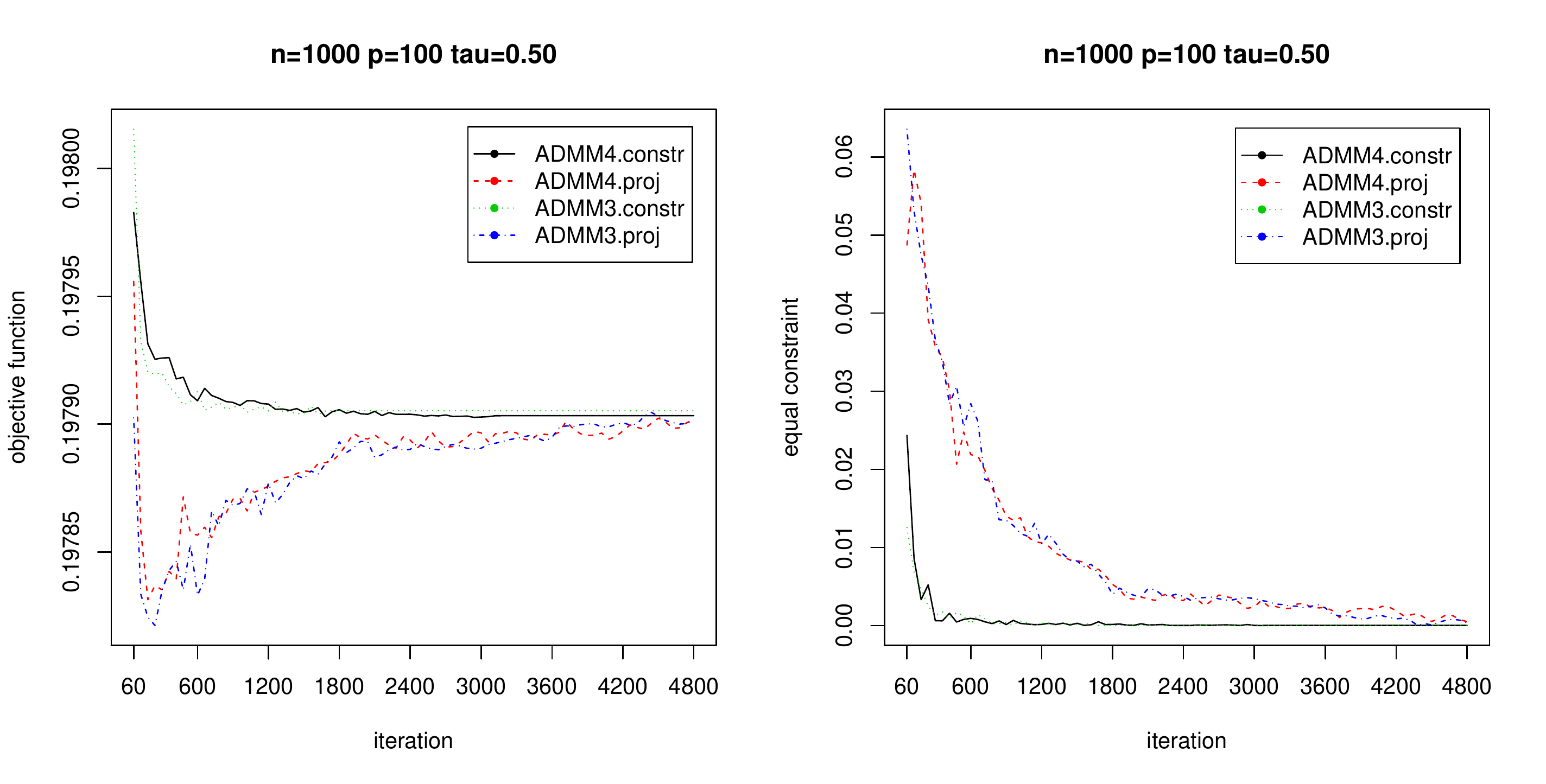}
\small
  \caption{ Objective function values and equal constraint values of four LCG-QR ADMM algorithms versus iteration steps for $n=1000 \ p=100$ and $\tau=0.50$.  }
\end{figure}
Despite the iteration steps of the algorithm, we also record the time in seconds for the whole computation process of ADMM.
To do a meaningful comparison, we make sure that all algorithms reach the same level of accuracy, that is, they obtain the same objective function and equal constraint values numerically. Table 1 reported the average computing time at three different quantile values $\tau$=0.25, 0.5, 0.75. It can be seen that ADMM4.Constr is the fastest and ADMM3.Proj is the slowest for all cases. In general, the ADMM4.Constr and ADMM4.Proj are faster than ADMM3.Constr and ADMM3.Proj. It is reasonable because the update for $\beta$ in former has a closed form but the latter is a lasso type problem which causes a nested-loop iteration. The ADMM4.Constr and ADMM3.Constr algorithms are faster than ADMM4.Proj and ADMM3.Proj algorithms because the update of $w$ in the latter needs a projection optimization.

Through these comparisons, we find that ADMM4.Constr is the most efficient among four algorithms, which coincide with the theoretical analysis in Section 3. Therefore, we only illustrate the statistical performance of ADMM4.constr in the remaining simulation studies.

\begin{table}[h!]
\caption{\small Average computing time of four ADMM algorithms for different data size. }
\footnotesize
\label{tab:1}
 \vspace{-0.3cm}
 \center
\begin{tabular}{ccccccc}
\hline
 & (n,p)  & $\tau$ & ADMM4.constr & ADMM4.proj & ADMM3.constr & ADMM3.proj \\
\hline
 & (500,50)& $0.25$ & 0.0783(0.0051)	& 0.3584(0.0743) &0.8175(0.0624)& 1.9725(0.4046)  \\  			
&  & $0.50$  &0.0732(0.0065)&0.4048(0.0677)&0.6312(0.0808)&1.3857(0.3199)\\  			
&  & $0.75$  & 0.0803(0.0070) &  0.3853(0.0929)&0.8822(0.0929)&2.0057(0.5939) \\  			
\hline
 & (1000,50) & $0.25$  & 0.1996(0.0163) &  0.6629(0.1296)& 1.1697(0.1157) & 2.5719(0.7473)  \\  			
&   & $0.50$  &0.1686(0.0139) &0.6459(0.1571) & 1.2498(0.1951)& 2.6929(0.5203)  \\  			
&   & $0.75$  &0.1735(0.0129) &0.5063(0.1269) & 1.2348(0.1478)& 2.3285(0.7249)\\
\hline
 & (1000,100) & $0.25$  &  0.4280(0.0233) & 0.8249(0.0905) & 1.8843(0.1225)& 5.2156(1.7852)\\
&   & $0.50$  & 0.3833(0.0349) &0.7516(0.1379) & 1.8039(0.1764)& 4.1530(1.4346) \\
&   & $0.75$  &0.4248(0.0198) & 0.9168(0.1837)& 2.0497(0.1775)  & 5.2461(1.4588) \\
\hline
\end{tabular}
\end{table}

\subsubsection{Performance of ADMM4.Constr}

In this subsection, we illustrate the finite-sample performance of the ADMM4.Constr algorithm.
Randomly generate $n$ training observations along with $n_{t}=2000$ test observations.
Recently, \cite{2020Generalized} explored the degrees of freedom of the linear constrained penalized quantile regression model. It is defined as $df_{\lambda}=|\calE|$, where $\calE=\{i:y_{i}-x_{i}^{T}\widehat{\beta}(\lambda)= 0, 1\leq i\leq n\}$ and $|\cdot|$ means the cardinality of one set. We use it to replace the nonzero number of parameter in model selection criterion of \cite{Peng2015iterative}. Define
\begin{align*}
\text{HBIC}(\lambda)=\text{log}\bigg(\sum_{i=1}^{n} \tilde\rho_{\tau}(y_{i} - x_{i}^{T}\widehat{\beta}(\lambda))\bigg)+df_{\lambda}\frac{\text{log}(\text{log} n)}{n}C_{n},
\end{align*}
where $\widehat{\beta}(\lambda)$ is the estimated value of $\beta$ for a given $\lambda$
and $C_{n}$ is a sequence of positive constants diverging to infinity as
$n$ increases. In practice, we recommend to take $C_{n}=O(\text{log}(p))$, which works well in a variety of settings. We select the value of $\lambda$ that minimizes HBIC($\lambda$).
 Using the optimal $\lambda$, we examine the quality of the proposed ADMM4.Constr algorithm from three aspects, namely, model selection accuracy, estimation accuracy and prediction accuracy.

The model selection accuracy is gauged by the average number of truly selected variables (Size), the percentage that $X_{1}$ was selected ($P_{1}$), the percentage that $X_{5}, X_{6}, X_{11}, X_{12}$ were selected ($P_{2}$).
The parameter $\beta$ is estimated on training data and the estimation accuracy is measured by
the absolute estimation error (AE) defined as
$ \text{AE}=\|\widehat{\beta}(\lambda)-\beta\|_{1}. $
To evaluate the goodness of fit, we calculate the mean absolute deviations (MAD) on test data defined as $\frac{1}{n_{t}}\sum\limits_{i=1}^{n_{t}}|x_{i}^{T}\beta-x_{i}^{T}\widehat{\beta}|$.
The prediction accuracy is evaluated by the mean absolute prediction error (MAPE) defined by
$ \text{MAPE}=\frac{1}{n_{t}}\sum\limits_{i=1}^{n_{t}}|y_{i}-x_{i}^{T}\widehat{\beta}|$ on test data.
The simulation results are summarized in Table 2.
We can see that the variables $X_{5}$, $X_{6}$, $X_{11}$, $X_{12}$ are selected for $\tau=0.25,0.5,0.75$ and variable $X_{1}$ is selected except for $\tau=0.5$, which is coincide with the true model in (5.1). This demonstrates the proposed ADMM4.Constr with the modified HBIC criterion is able to select the true model.
Moreover, the AE becomes larger with the increase of $p$ for the same sample size $n$ and becomes smaller with the increase of $n$ for the same dimension $p$.

 \begin{table}[h!]
\caption{\small Performance of ADMM4.Constr for normal error model with different data sizes. }
\footnotesize
\label{tab:1}
 \vspace{-0.3cm}
 \center
\begin{tabular}{ccccccccc}
\hline
 & (n,p)  & $\tau$ & Size & $P_{1}$ &  $P_{2}$ & AE & MAD  & MAPE  \\
\hline
 & (1000,50)& $0.25$ & 5(0)	& 	1 & 1& 0.2122(0.0553) &0.0700(0.0235)& 0.4627(0.0141)  \\  				
&  & $0.50$  &4(0) & 0&	1& 0.0901(0.0368) &  0.0245(0.0079) &0.3969(0.0094)	\\  					
&  & $0.75$  & 5(0)  & 1 & 1 &0.2168(0.0552)  & 0.0717(0.0235) &0.4539(0.0145)\\ 			
\hline
 & (1000,100) & $0.25$  & 5(0) & 1 & 1 &0.2711(0.0622)  &  0.0792(0.0243) & 0.4568(0.0143)\\  	
&   & $0.50$  & 4(0) &0  & 1 & 0.0992(0.0271) & 0.0208(0.0059)  &0.4011(0.0079)\\
&   & $0.75$  & 5(0) & 1 & 1 &0.2445(0.0497)  &0.0789(0.0215) &0.4519(0.0113)\\ 		
\hline
 & (2000,100) & $0.25$  & 5(0) & 1 & 1 &0.1796(0.0436)  &0.0437(0.0146)&0.4581(0.0100)\\ 	
&   & $0.50$  & 4(0) & 0 & 1 &0.0918(0.0249)  &0.0173(0.0043)  & 0.3963(0.0078)\\  			
&   & $0.75$  & 5(0) & 1 & 1 &0.1762(0.0393) &0.0448(0.0150) & 0.4679(0.0129)\\ 	
\hline
\end{tabular}
\end{table}

\subsection{Simulation for parallel LCG-QR ADMM}

In this section, we investigate the performance of parallel ADMM4.Constr algorithm for large scale data. The data here are generated in the same way as those in section 5.1 with $(N,p)=(50000,50)$. The data are randomly and evenly split into $M$ subsets of size $N/M$ to mimic the distributed environment. For the number of partitions or say local machines, we consider $M= 20, 40, 50, 80, 100, 125, 200$.

Figure 4 shows the estimation accuracy and computation efficiency versus number of partitions $M$ at
third quartile $\tau=0.75$. The left panel shows the absolute estimation error (AE) and the right panel shows the computing time (Time) of the algorithm. We can see slight increase in the estimation error as the number of partitions $M$ increases, and the estimation error towards stability in the end. Notice that the total sample size $N$ is fixed and then the local sample size $n$ is decreasing as $M$ increases. The local estimator $\widehat{\beta}^{k}$ becomes less accurate as the sample size $n$ in each partition
reduces. Moreover, the update for $\beta$ is a combination of $\widehat{\beta}^{k}$, so the estimation error of $\beta$ increases as $M$ increases. The computing time of our method decreases obviously as $M$ increase and becomes stable finally. This is easy to understand that smaller sample size in each partition leads to a faster running speed. Thus, the parallel ADMM4.Constr remarkably reduce the
computation time, which is critical to big data analysis.

\begin{figure}[ht]
\centering
{
\includegraphics[width=7.5cm,height=7.5cm]{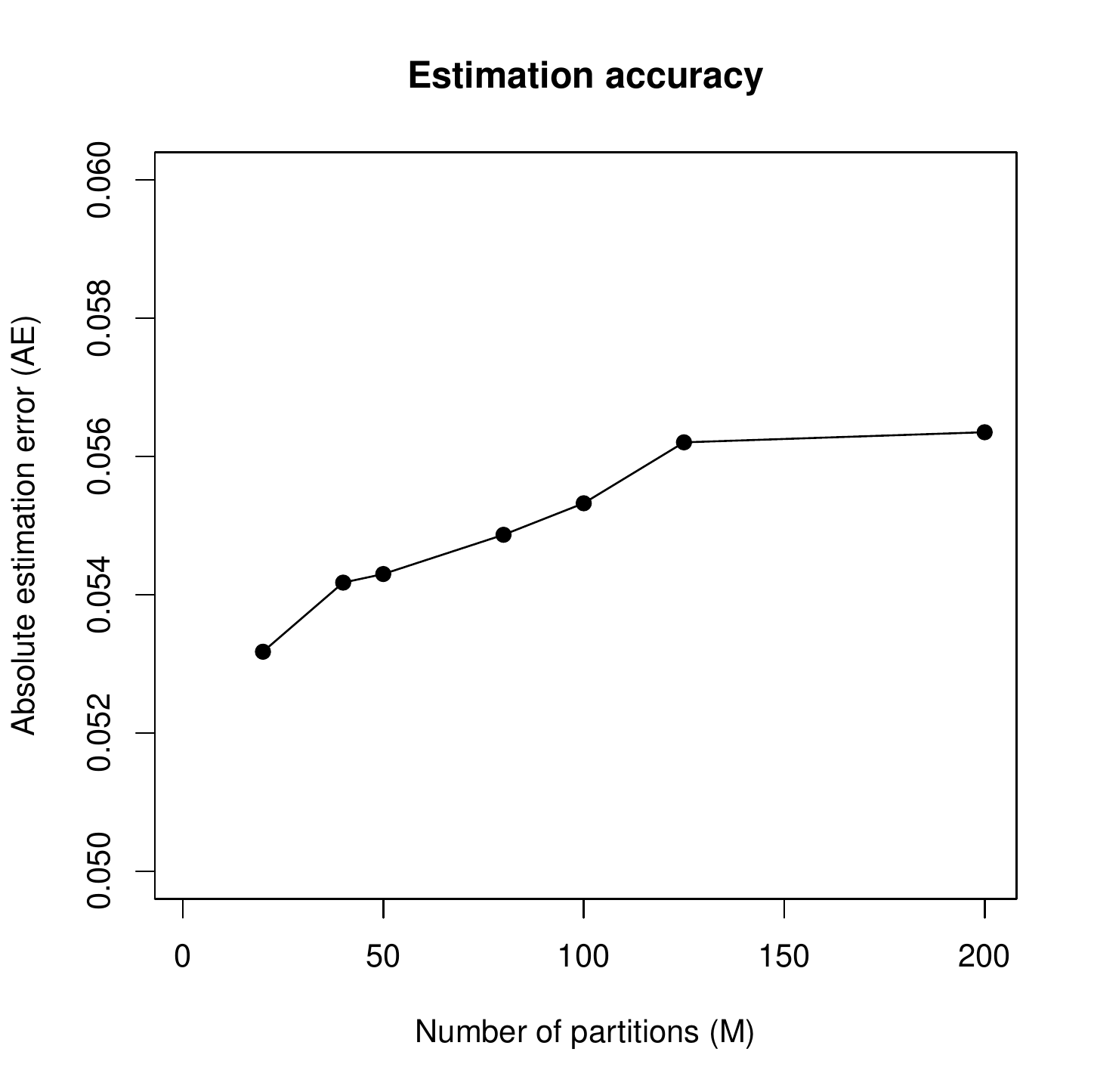}}
\hspace{0in}
{
\includegraphics[width=7.5cm,height=7.5cm]{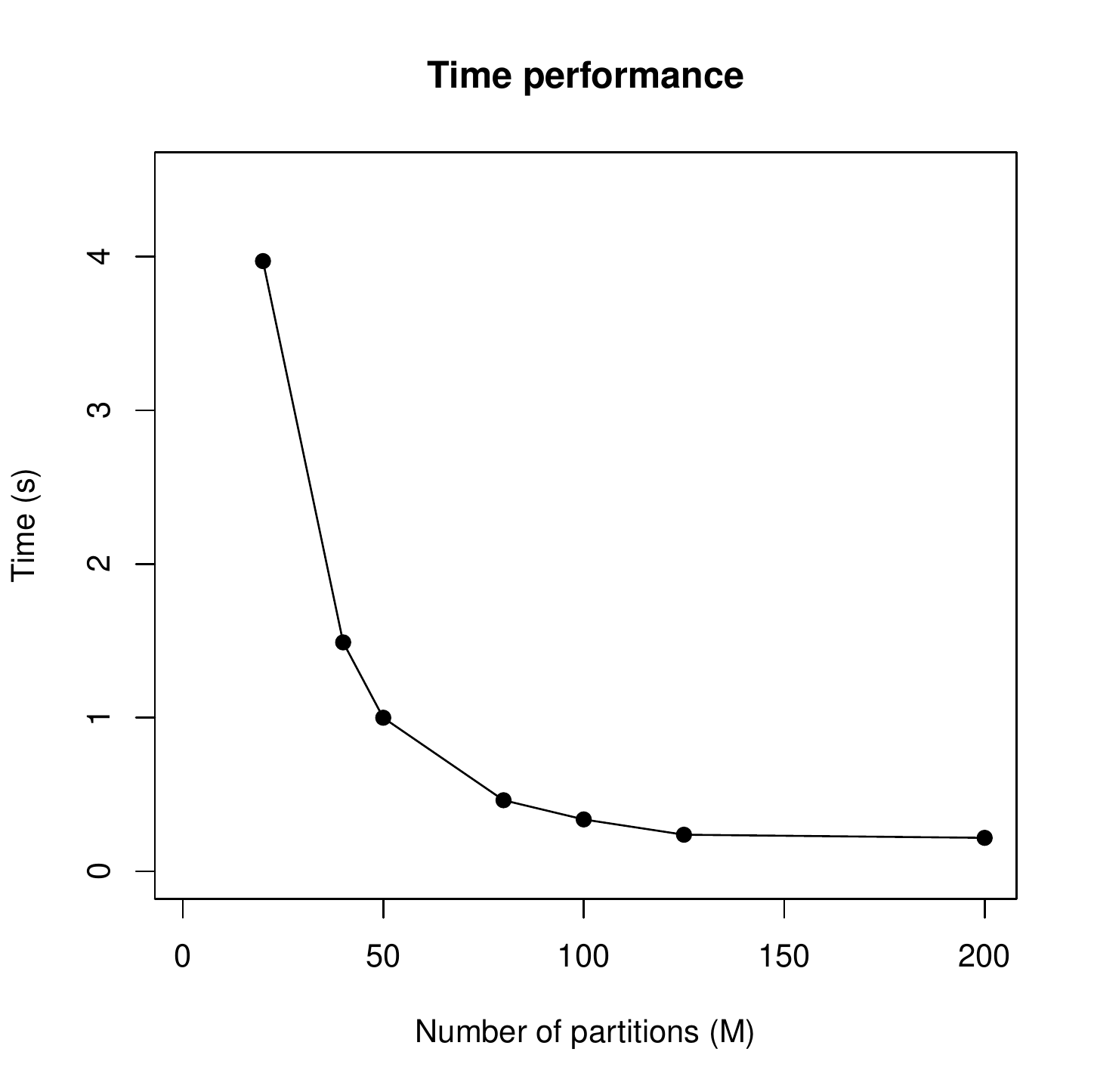}}
\caption{The estimation error and computational time of parallel ADMM4.Constr for different $M$ with $\tau=0.75$.}
\end{figure}

\section{Real data analysis}

In this section, we apply the proposed parallel ADMM4.Constraint algorithm to a real data set: greenhouse gas (GHG) observing network data set reported by the UCI machine learning repository. This massive data set consists of 954,840 samples observed within three months of 2010. The response variable is GHG concentrations of synthetic observations
and the predictors are GHG concentrations of tracers emitted from 15 spatial regions (denoted as Reg1-Reg15), which has a positive relationship between them.
The concentrations of tracers in some regions may not affect the concentrations of synthetic. Therefore, the penalized quantile regression model with positive constraints of coefficients is suitable here to identify the effective tracers.

To evaluate the performance of the algorithm, we divide the full data into training data and
testing data equally, both of them have 477,420 observations. We randomly split the training data set into $M=10, 20, 50, 80, 100, 200$ subsets in order to mimic distributed computing. The sample size of first $M-1$ subsets is $\lfloor \frac{477,420}{M}\rfloor$ and the last subset contains all the remaining data, where $\lfloor a\rfloor$ denotes the largest integer less than $a$. Then we estimate the coefficients using the training data and calculate the mean of absolute prediction error MAPE defined in section 5.1.2 based on the test data. We also report the average number of
selected nonzero coefficients (Size) and the running time (Time) of the algorithm.
To make a comparison between the global algorithm and the parallel algorithm,
we fitted the model with the whole training data and test data, that is $M=1$.
All these results are computed at $\tau=0.5$ and illustrated in Table 3.

 \begin{table}[h!]

\caption{\small Performance of parallel ADMM4.Constr algorithm for the GHG data set. }
\footnotesize
\label{tab:1}
 \vspace{-0.3cm}
 \center
\begin{tabular}{cccccccc}
\hline
& M=1  & M=10 & M=20 &  M=50 & M=80   & M=100  & M=200\\
\hline
 $\text{MAPE}$ & 13.9005 & 13.9256 & 13.9196 & 13.9293  &13.9167  &  13.9231&  13.9167   \\  				
\hline
 $\text{Size}$ &  12 &12 & 12 &	12&   12&  12&	12\\  						
 \hline
 $\text{Time}$ &  16.2800  & 12.9170   &  6.7785  & 1.7860   &0.9526  &0.8419  & 0.5268 \\		
\hline
\end{tabular}
\end{table}

It can be seen that the MAPE for different number of $M$ from 10 to 200 are always similar to each other, and very close to that of the global regression method ($M=1$). It shows the algorithm runs faster as $M$ increases because larger $M$ leads to smaller sample size in each subset and thus increase the computation speed. Through the comparison of running time, we found that the parallel method is much more time-efficient for large-scale data.
Besides, the variables Reg1, Reg5 and Reg6 are not selected in the model by all settings of $M$, which informs these regions have almost no effect on GHG concentrations of synthetic.
 All of these illustrates the excellent performance of our method in the real data analysis.

\section{Discussions and Future Work}
In this paper, we consider the computation method based on ADMM algorithms of general penalized quantile regression with linear constraints for big data. The key idea of our method is to transform the optimization problem (\ref{lcgqproblem}) into ADMM form by introducing new variables to represent the loss function, general penalty and linear constraints. We first extended the classic ADMM to the case of more than three blocks and then apply it to LCG-QR model.
The most efficient ADMM4.Constr algorithm has explicit expression of each  parameter during the update schedule. We further propose the parallel ADMM algorithm for large-sample data, which can be computed distributely on each machine. It runs faster than the global method and has almost the same level of estimation accuracy as the global method. We conduct numerical studies to demonstrate the efficiency of the proposed ADMM and parallel ADMM algorithms.

 Actually, there are some alternatives for the separate optimization program with $l\geq 3$ variables and functions. For example, \cite{2012ALTERNATING} combined the ADMM with a Gaussian back substitution procedure and prove its convergence. \cite{2015ONFULL} consider a Jacobian scheme of the augmented Lagrangian method and \cite{Deng2017Parallel} added a proximal term on the Jacobian ADMM. In the future, we would like to explore these optional transformations of Lagrangian algorithms to solve the LCG-QR model for big data, both in large sample-size and extra high dimension.

\section*{Appendix A}
\subsection*{Proof of Theorem 1}
We take the direct extension of ADMM with four variables as an example to show the sufficient conditions ensuring the convergence of (2.5). The idea is to show that the variable $x_{1},x_{2},x_{3},x_{4}$ can be regarded as two parts of variables, which reduces to a special case of the classic ADMM.

According to the first-order optimality conditions of the minimization problems in (2.5), we have
\begin{align}
\left\{
\begin{aligned}
&f_{1}(x_{1})-f_{1}(x_{1}^{k+1})+\gamma(x_{1}-x_{1}^{k+1})^{T}A_{1}^{T}
(A_{1}x_{1}^{k+1}+A_{2}x_{2}^{k}+A_{3}x_{3}^{k}+A_{4}x_{4}^{k}-c+u^{k})\geq 0, \\
&f_{2}(x_{2})-f_{2}(x_{2}^{k+1})+\gamma(x_{2}-x_{2}^{k+1})^{T}A_{2}^{T}
(A_{1}x_{1}^{k+1}+A_{2}x_{2}^{k+1}+A_{3}x_{3}^{k}+A_{4}x_{4}^{k}-c+u^{k})\geq 0, \\
&f_{3}(x_{3})-f_{3}(x_{3}^{k+1})+\gamma(x_{3}-x_{3}^{k+1})^{T}A_{3}^{T}
(A_{1}x_{1}^{k+1}+A_{2}x_{2}^{k+1}+A_{3}x_{3}^{k+1}+A_{4}x_{4}^{k}-c+u^{k})\geq 0,\\
&f{4}(x_{4})-f_{4}(x_{4}^{k+1})+\gamma(x_{4}-x_{4}^{k+1})^{T}A_{4}^{T}
(A_{1}x_{1}^{k+1}+A_{2}x_{2}^{k+1}+A_{3}x_{3}^{k+1}+A_{4}x_{4}^{k+1}-c+u^{k})\geq 0.
\end{aligned}
\right. \tag{A1.1}
\end{align}

Firstly, assume $A_{2}^{T}A_{3}=0$, $A_{2}^{T}A_{4}=0$, $A_{3}^{T}A_{4}=0$.
Then, it follows from (A1.1) that
\begin{align}
\left\{
\begin{aligned}
&f_{1}(x_{1})-f_{1}(x_{1}^{k+1})+\gamma(x_{1}-x_{1}^{k+1})^{T}A_{1}^{T}
(A_{1}x_{1}^{k+1}+A_{2}x_{2}^{k}+A_{3}x_{3}^{k}+A_{4}x_{4}^{k}-c+u^{k})\geq 0, \\
&f_{2}(x_{2})-f_{2}(x_{2}^{k+1})+\gamma(x_{2}-x_{2}^{k+1})^{T}A_{2}^{T}
(A_{1}x_{1}^{k+1}+A_{2}x_{2}^{k+1}-c+u^{k})\geq 0, \\
&f_{3}(x_{3})-f_{3}(x_{3}^{k+1})+\gamma(x_{3}-x_{3}^{k+1})^{T}A_{3}^{T}
(A_{1}x_{1}^{k+1}+A_{3}x_{3}^{k+1}-c+u^{k})\geq 0,\\
&f{4}(x_{4})-f_{4}(x_{4}^{k+1})+\gamma(x_{4}-x_{4}^{k+1})^{T}A_{4}^{T}
(A_{1}x_{1}^{k+1}+A_{4}x_{4}^{k+1}-c+u^{k})\geq 0.
\end{aligned}
\right. \tag{A1.2}
\end{align}
which is also the first-order optimality condition of the scheme
\begin{align}
\left\{
\begin{aligned}
x_{1}^{k+1}&=\arg\min_{x_{1}}\{ f_{1}(x_{1})+\frac{\gamma}{2}\|A_{1}x_{1}+A_{2}x_{2}^{k}
+A_{3}x_{3}^{k}+A_{4}x_{4}^{k}-c+u\|^{2}\},\\
(x_{2}^{k+1},x_{3}^{k+1},x_{4}^{k+1})&=\arg\min_{x_{2},x_{3},x_{4}}\{ f_{2}(x_{2})+f_{3}(x_{3})+f_{4}(x_{4})+\frac{\gamma}{2}\|A_{1}x_{1}^{k+1}+A_{2}x_{2}
+A_{3}x_{3}+A_{4}x_{4}-c+u\|^{2}\},\\
u^{k+1} &= u^{k}+(A_{1}x_{1}^{k+1}+A_{2}x_{2}^{k+1}+A_{3}x_{3}^{k+1}+A_{4}x_{4}^{k+1}-c).
\end{aligned}
\right.  \tag{A1.3}
\end{align}
Clearly, (A1.3) is a specific application of the classic ADMM by
regarding ($x_{2}$, $x_{3}$, $x_{4}$) as one variable.
Thus, existing convergence results
for the classic ADMM hold for the special case of (2.5)
with the orthogonality condition $A_{2}^{T}A_{3}=0$ and $A_{2}^{T}A_{4}=0$ and $A_{3}^{T}A_{4}=0$.

Similarly, assume $A_{1}^{T}A_{2}=0$ and $A_{3}^{T}A_{4}=0$ , it follows from (A1.1) that
\begin{align}
\left\{
\begin{aligned}
&f_{1}(x_{1})-f_{1}(x_{1}^{k+1})+\gamma(x_{1}-x_{1}^{k+1})^{T}A_{1}^{T}
(A_{1}x_{1}^{k+1}+A_{3}x_{3}^{k}+A_{4}x_{4}^{k}-c+u^{k})\geq 0, \\
&f{2}(x_{2})-f_{2}(x_{2}^{k+1})+\gamma(x_{2}-x_{2}^{k+1})^{T}A_{2}^{T}
(A_{2}x_{2}^{k+1}+A_{3}x_{3}^{k}+A_{4}x_{4}^{k}-c+u^{k})\geq 0, \\
&f_{3}(x_{3})-f_{3}(x_{3}^{k+1})+\gamma(x_{3}-x_{3}^{k+1})^{T}A_{3}^{T}
(A_{1}x_{1}^{k+1}+A_{2}x_{2}^{k+1}+A_{3}x_{3}^{k+1}-c+u^{k})\geq 0, \\
&f_{4}(x_{4})-f_{4}(x_{4}^{k+1})+\gamma(x_{4}-x_{4}^{k+1})^{T}A_{4}^{T}
(A_{1}x_{1}^{k+1}+A_{2}x_{2}^{k+1}+A_{4}x_{4}^{k+1}-c+u^{k})\geq 0,
\end{aligned}
\right. \tag{A1.4}
\end{align}
which is also the first-order optimality condition of the scheme
\begin{align}
\left\{
\begin{aligned}
(x_{1}^{k+1},x_{2}^{k+1})&=\arg\min_{x_{1},x_{2}}\{ f_{1}(x_{1})+f_{2}(x_{2})
+\frac{\gamma}{2}\|A_{1}x_{1}+A_{2}x_{2}+A_{3}x_{3}^{k}+A_{4}x_{4}^{k}-c+u\|^{2}\},\\
(x_{3}^{k+1},x_{4}^{k+1})&=\arg\min_{x_{3},x_{4}}\{ f_{3}(x_{3})+f_{4}(x_{4})
+\frac{\gamma}{2}\|A_{1}x_{1}^{k+1}+A_{2}x_{2}^{k+1}
+A_{3}x_{3}+A_{4}x_{4}-c+u\|^{2}\},\\
u^{k+1}&= u^{k}+u(A_{1}x_{1}^{k+1}+A_{2}x_{2}^{k+1}+A_{3}x_{3}^{k+1}+A_{4}x_{4}^{k+1}-c).
\end{aligned}
\right.  \tag{A1.5}
\end{align}
Clearly, (A1.5) is a specific application of the classic ADMM by
regarding ($x_{1}$, $x_{2}$) as one variable
 and regarding ($x_{3}$, $x_{4}$) as one variable.
Existing convergence results for the classic ADMM hold for the special case of (2.5)
with the orthogonality condition $A_{1}^{T}A_{2}=0$ and $A_{3}^{T}A_{4}=0$.

Similarly, assume $A_{1}^{T}A_{2}=0$ and $A_{1}^{T}A_{3}=0$ and $A_{2}^{T}A_{3}=0$, it follows from (A1.1) that
\begin{align}
\left\{
\begin{aligned}
&f_{1}(x_{1})-f_{1}(x_{1}^{k+1})+\gamma(x_{1}-x_{1}^{k+1})^{T}A_{1}^{T}
(A_{1}x_{1}^{k+1}+A_{4}x_{4}^{k}-c+u^{k})\geq 0, \\
&f_{2}(x_{2})-f_{2}(x_{2}^{k+1})+\gamma(x_{2}-x_{2}^{k+1})^{T}A_{2}^{T}
(A_{2}x_{2}^{k+1}+A_{4}x_{4}^{k}-c+u^{k})\geq 0, \\
&f_{3}(x_{3})-f_{3}(x_{3}^{k+1})+\gamma(x_{3}-x_{3}^{k+1})^{T}A_{3}^{T}
(A_{3}x_{3}^{k+1}+A_{4}x_{4}^{k}-c+u^{k})\geq 0,\\
&f_{4}(x_{4})-f_{4}(x_{4}^{k+1})+\gamma(x_{4}-x_{4}^{k+1})^{T}A_{4}^{T}
(A_{1}x_{1}^{k+1}+A_{2}x_{2}^{k+1}+A_{3}x_{3}^{k+1}+A_{4}x_{4}^{k+1}-c+u^{k})\geq 0,
\end{aligned}
\right. \tag{A1.6}
\end{align}
which is also the first-order optimality condition of the scheme
\begin{align}
\left\{
\begin{aligned}
(x_{1}^{k+1},x_{2}^{k+1},x_{3}^{k+1})&=\arg\min_{x_{1},x_{2},x_{3}}\{ f_{1}(x_{1})+f_{2}(x_{2})+f_{3}(x_{3})+\frac{\gamma}{2}\|A_{1}x_{1}+A_{2}x_{2}+A_{3}x_{3}
+A_{4}x_{4}^{k}-c+u\|^{2}\},\\
x_{4}^{k+1}&=\arg\min_{x_{4}}\{ f_{4}(x_{4})+\frac{\gamma}{2}\|A_{1}x_{1}^{k+1}+A_{2}x_{2}^{k+1}
+A_{3}x_{3}^{k+1}+A_{4}x_{4}-c+u\|^{2}\},\\
u^{k+1}& = u^{k}+(A_{1}x_{1}^{k+1}+A_{2}x_{2}^{k+1}+A_{3}x_{3}^{k+1}+A_{4}x_{4}^{k+1}-c).
\end{aligned}
\right.  \tag{A1.7}
\end{align}

Clearly, (A1.7) is a specific application of the classic ADMM by
regarding ($x_{1}$, $x_{2}$, $x_{3}$) as one variable.
Thus, existing convergence results for the classic ADMM hold for the special case of (2.5)
with the orthogonality condition $A_{1}^{T}A_{2}=0$ and $A_{1}^{T}A_{3}=0$ and $A_{2}^{T}A_{3}=0$.

Therefore, the conditions in Theorem 1 ensure that the direct extended ADMM with $N$ variables $x_{1},\ldots,x_{N}$ has the same first-order optimality condition as the classic ADMM with two blocks of variables $(x_{1},\ldots,x_{M})$ and $(x_{M+1},\ldots,x_{N})$ for some constant $M$. Then the convergence of direct extension in (2.5) is implied by the well known results in the existing ADMM literature.

\subsection*{Proof of residuals in Section 2.3 }

According to the direct extension of ADMM in Section 2.2, the augmented Lagrangian for problem (2.7) is
\begin{align*}
\mathcal{L}_{\gamma}(x_{1},x_{2},x_{3},x_{4},u)=&\sum_{l=1}^{4} f_{l}(x_{l})
+\frac{\gamma}{2}\|A_{1}x_{1}+A_{2}x_{2}+A_{3}x_{3}+A_{4}x_{4}-c+u\|^{2}.
\end{align*}
The updating scheme is
\begin{align}
x_{1}^{k+1} &= \arg\min_{x_{1}}\mathcal{L}_{\gamma}(x_{1},x_{2}^{k},x_{3}^{k},x_{4}^{k},u^{k}),\tag{A2.1}\\
x_{2}^{k+1} &= \arg\min_{x_{2}}\mathcal{L}_{\gamma}(x_{1}^{k+1},x_{2},x_{3}^{k},x_{4}^{k},u^{k}),\tag{A2.2}\\
x_{3}^{k+1} &= \arg\min_{x_{3}}\mathcal{L}_{\gamma}(x_{1}^{k+1},x_{2}^{k+1},x_{3},x_{4}^{k},u^{k}),\tag{A2.3}\\
x_{4}^{k+1} &= \arg\min_{x_{4}}\mathcal{L}_{\gamma}(x_{1}^{k+1},x_{2}^{k+1},x_{3}^{k+1},x_{4},\eta^{k}),\tag{A2.4}\\
u^{k+1} &= u^{k}+(A_{1}x_{1}^{k+1}+A_{2}x_{2}^{k+1}+A_{3}x_{3}^{k+1}+A_{4}x_{4}^{k+1}-c).\tag{A2.5}
\end{align}
The necessary and sufficient optimality conditions for the problem (2.7)
are primal feasibility,
\begin{align}
A_{1}x^{*}_{1}+A_{2}x^{*}_{2}+A_{3}x^{*}_{3}+A_{4}x^{*}_{4}-c=0, \tag{A2.6}
\end{align}
and dual feasibility,
\begin{align}
0 \in \partial f_{1}(x^{*}_{1})+\gamma A_{1}^{T}u^{*}, \tag{A2.7}\\
0 \in \partial f_{2}(x^{*}_{2})+\gamma A_{2}^{T}u^{*}, \tag{A2.8}\\
0 \in \partial f_{3}(x^{*}_{3})+\gamma A_{3}^{T}u^{*}, \tag{A2.9}\\
0 \in \partial f_{4}(x^{*}_{4})+\gamma A_{4}^{T}u^{*}. \tag{A2.10}
\end{align}
The updating scheme is stopped when the primal and dual residuals are small enough.
 We tend to demonstrate the solution derived from (A2.1-A2.5) satisfy the above optimality conditions with the proposed stopping criterion.

According to the primal feasibility (A2.6), we take the primal residual $r$ at iteration step $k$ as
\begin{align*}
r^{k} = A_{1}v^{k}_{1}+A_{2}v^{k}_{2}+A_{3}v^{k}_{3}+A_{4}v^{k}_{4}-c.
\end{align*}
Due to the fact that $x_{4}^{k+1}$ minimize $\mathcal{L}_{\gamma}(x_{1}^{k+1},x_{2}^{k+1},x_{3}^{k+1},x_{4},u^{k})$ in (A2.4), we have
\begin{align*}
0 &\in \partial f_{4}(x^{k+1}_{4})+\gamma A_{4}^{T}(A_{1}x_{1}^{k+1}+A_{2}x_{2}^{k+1}+A_{3}x_{3}^{k+1}+A_{4}x_{4}^{k+1}-c+u^{k})\\
&=\partial f_{4}(x^{k+1}_{4})+\gamma A_{4}^{T}(r^{k+1}+u^{k})\\
&=\partial f_{4}(x^{k+1}_{4})+\gamma A_{4}^{T}u^{k+1}.
\end{align*}
This shows that $x^{k+1}_{4}$ and $u^{k+1}$ satisfy the dual feasibility (A2.10).

Also, due to the fact that $x_{3}^{k+1}$ minimize $\mathcal{L}_{\gamma}(x_{1}^{k+1},x_{2}^{k+1},x_{3},x_{4}^{k},u^{k})$ in (A2.3), we have
\begin{align*}
0 &\in \partial f_{3}(x^{k+1}_{3})+\gamma A_{3}^{T}(A_{1}x_{1}^{k+1}+A_{2}x_{2}^{k+1}+A_{3}x_{3}^{k+1}+A_{4}x_{4}^{k}-c+u^{k})\\
&= \partial f_{3}(x^{k+1}_{3})+\gamma A_{3}^{T}(A_{1}x_{1}^{k+1}+A_{2}x_{2}^{k+1}+A_{3}x_{3}^{k+1}+A_{4}x_{4}^{k+1}
-c+A_{4}x_{4}^{k}-A_{4}x_{4}^{k+1}+u^{k})\\
&= \partial f_{3}(x^{k+1}_{3})+\gamma A_{3}^{T}(r^{k+1}+A_{4}(x_{4}^{k}-x_{4}^{k+1})+u^{k})\\
&= \partial f_{3}(x^{k+1}_{3})+\gamma A_{3}^{T}u^{k+1}+\gamma A_{3}^{T}A_{4}(x_{4}^{k}-x_{4}^{k+1}).
\end{align*}
The variable $x_{3}^{k+1}$ and $u^{k+1}$ would satisfy (A2.9) only when the last term in the above formula is close to zero. Thus, we take
\begin{align*}
s_{1}^{k+1}=\gamma A_{3}^{T}A_{4}(x_{4}^{k+1}-x_{4}^{k})
\end{align*}
 as a dual residual.

Similarly, due to the fact that $x_{2}^{k+1}$ minimize $\mathcal{L}_{\gamma}(x_{1}^{k+1},x_{2},x_{3}^{k},x_{4}^{k},u^{k})$ in (A2.2), we have
\begin{align*}
0 &\in \partial f_{2}(x^{k+1}_{2})+\gamma A_{2}^{T}(A_{1}v_{1}^{k+1}+A_{2}v_{2}^{k+1}+A_{3}v_{3}^{k}+A_{4}v_{4}^{k}-c+u^{k})\\
&=\partial f_{2}(x^{k+1}_{2})+\gamma A_{2}^{T}(r^{k+1}+A_{3}(x_{3}^{k}-x_{3}^{k+1})+A_{4}(x_{4}^{k}-x_{4}^{k+1})+u^{k})\\
&=\partial f_{2}(x^{k+1}_{2})+ A_{2}^{T}u^{k+1}+\gamma A_{2}^{T}A_{3}(x_{3}^{k}-x_{3}^{k+1})+\gamma A_{2}^{T}A_{4}(x_{4}^{k}-x_{4}^{k+1}).
\end{align*}
The variable $x_{2}^{k+1}$ and $u^{k+1}$ would satisfy (A2.8) only when the sum of last two terms in the above formula are close to zero. Thus, we take
\begin{align*}
s_{2}^{k+1}=\gamma A_{2}^{T}A_{3}(v_{3}^{k+1}-v_{3}^{k})+\gamma A_{2}^{T}A_{4}(v_{4}^{k+1}-v_{4}^{k})
\end{align*}
 as a  dual residual.

Finally, due to the fact that $x_{1}^{k+1}$ minimize $\mathcal{L}_{\gamma}(x_{1},x_{2}^{k},x_{3}^{k},x_{4}^{k},u^{k})$ in (A2.1), we have
\begin{align*}
0 &\in \partial f_{1}(x^{k+1}_{1})+\gamma A_{1}^{T}(A_{1}x_{1}^{k+1}+A_{2}x_{2}^{k}+A_{3}x_{3}^{k}+A_{4}x_{4}^{k}-c+u^{k})\\
&= \partial f_{1}(x^{k+1}_{1})+\gamma A_{1}^{T}(r^{k+1}
+A_{2}(v_{2}^{k}-v_{2}^{k+1})+A_{3}(v_{3}^{k}-v_{3}^{k+1})+A_{4}(v_{4}^{k}-v_{4}^{k+1})+u^{k})\\
&= \partial f_{1}(x^{k+1}_{1})+A_{1}^{T}u^{k+1}+\gamma A_{1}^{T}
A_{2}(x_{2}^{k}-x_{2}^{k+1})+\gamma A_{1}^{T}A_{3}(x_{3}^{k}-x_{3}^{k+1})+\gamma A_{1}^{T}A_{4}(x_{4}^{k}-x_{4}^{k+1}).
\end{align*}
The variable $x_{1}^{k+1}$ and $u^{k+1}$ would satisfy (A2.7) only when the sum of last three terms in the above formula are close to zero. Thus, we take
\begin{align*}
s_{3}^{k+1}=\gamma A_{1}^{T}A_{2}(x_{2}^{k+1}-x_{2}^{k})+\gamma A_{1}^{T}A_{3}(x_{3}^{k+1}-x_{3}^{k})
+\gamma A_{1}^{T}A_{4}(x_{4}^{k+1}-x_{4}^{k})
\end{align*}
 as a dual residual.

 To this end, we obtain the primal residual $r$ and dual residuals $s_{1}, s_{2}, s_{3}$ to ensure the updating variable $(x_{1},x_{2},x_{3},x_{4},u)$ at the last iteration step satisfy the primal and dual feasibility, which leads to the optimal solution of the problem (2.7).

\section*{ Appendix B }
\subsection*{ The other three LCG-QR ADMM algorithms}
The ADMM4.Proj, ADMM3.Constr and ADMM3.Proj algorithms with Lasso penalty are shown here, the solutions with MCP and SCAD penalty are postponed in Appendix C.
\subsubsection*{ADMM4.Proj}
By introducing new variables
$r \in \mathbb{R}^{n}$, $z \in \mathbb{R}^{m}$ and $w \in \mathbb{R}^{p}$, we formulate the problem (1.1) as
\begin{align*}
& \rho_{\tau}(r) + \lambda\|D\beta\|_{1} + \phi(w) \nonumber \\
&\text{subject to \
  $y - X\beta = r,$ \
  $z = D\beta,$ \
  $\beta = w,$}
\end{align*}
where $\phi(w) = 0$ if $Cw \geq d$ and $Ew = f$ and $= \infty$ otherwise. Denote \begin{align*}
A_{1}=
 \begin{pmatrix}
 X \\
 D \\
 I_{p}
 \end{pmatrix}, \quad
A_{2}
 = \begin{pmatrix}
  I_{n} \\
  0_{m} \\
  0_{p}
 \end{pmatrix},\quad
 A_{3}
 = \begin{pmatrix}
  0_{n} \\
  -I_{m} \\
  0_{p}
 \end{pmatrix},\quad
 A_{4}
 = \begin{pmatrix}
  0_{n} \\
  0_{m} \\
  -I_{p}
 \end{pmatrix},\quad
c=\begin{pmatrix}
  y\\
  0_{m} \\
  0_{p}
 \end{pmatrix}.
\end{align*}The augmented Lagrangian
is
\begin{align*}
\mathcal{L}_{\gamma}(\beta,r,z,w,u)= \rho_{\tau}(r) +  \lambda\|D\beta\|_{1} +\phi(w)+\frac{\gamma}{2}\|A_{1}\beta+A_{2}r+A_{3}z+A_{4}w-c+u\|_{2}^{2},
\end{align*}
where $u=(u_{1}^{T},u_{2}^{T},u_{3}^{T})^{T}$ is the scaled dual variable satisfying $u_{1} \in \mathbb{R}^{n}$,
$u_{2} \in \mathbb{R}^{m}$, $u_{3} \in \mathbb{R}^{p}$.
The ADMM updating rules are as follows,
\begin{align*}
\beta^{(k+1)}
&=(X^{T}X+D^{T}D+I_{p})^{-1} \big( X^{T}( y-r^{(k)}-u_{1}^{(k)})+ D^{T}(z^{(k)}-u_{2}^{(k)})
+ w^{(k)}-u_{3}^{(k)}\big),\\
r_{i}^{(k+1)}
&=[y_{i}-x_{i}^{T}\beta^{(k+1)}-u_{1i}^{(k)}-\tau/\gamma]_{+}-
[-y_{i}+x_{i}^{T}\beta^{(k+1)}+u_{1i}^{(k)}+(\tau-1)/\gamma]_{+},\\
z_{j}^{(k+1)}
 &=(D_{j}\beta^{(k+1)}+u_{2j}^{(k)}-\lambda/\gamma)_{+}
 -(-D_{j}\beta^{(k+1)}-u_{2j}^{(k)}-\lambda/\gamma)_{+},\\
w^{(k+1)}&=\text{Proj$_{\mathcal{C}}$}(\beta^{(k+1)}+u_{3}^{(k)}),\\
u^{(k+1)}&=u^{(k)}+
\begin{pmatrix}
 X\beta^{(k+1)}+r^{(k+1)}-y \\
 D\beta^{(k+1)}-z^{(k+1)} \\
 \beta^{(k+1)}-w^{(k+1)}
\end{pmatrix}.
\end{align*}
The update for $w$ is a projection on set $\mathcal{C}=\{w: Cw\geq d, Ew=f \}$.
The stopping criterion of algorithm ADMM4.Proj follows similar way as ADMM4.Constr.

\subsubsection*{ADMM3.Constr}
By introducing new variables $r \in \mathbb{R}^{n}$ and $w \in \mathbb{R}^{q}$, we formulate the problem (1.1) as
\begin{align*}
& \rho_{\tau}(r) + \lambda\|D\beta\|_1 + \phi(w)\nonumber \\
&\text{subject to \
  $y - X\beta = r,$ \
  $C\beta - w = d,$ \
  $E\beta = f,$ }
\end{align*}
where $\phi(w) = 0$ if $w \geq 0$ componentwise and $= \infty$ otherwise.
Denote \begin{align*}
A_{1}=
 \begin{pmatrix}
 X \\
 C \\
  E
 \end{pmatrix}, \quad
A_{2}
 = \begin{pmatrix}
  I_{n} \\
  0_{q} \\
  0_{s}
 \end{pmatrix},\quad
A_{3}
 = \begin{pmatrix}
 0_{n} \\
  -I_{q} \\
  0_{s}
 \end{pmatrix}\quad
c=\begin{pmatrix}
  y\\
  d \\
  f
 \end{pmatrix}.
\end{align*}
The augmented Lagrangian is
\begin{align*}
\mathcal{L}_{\gamma}(\beta,r,w,u)&= \rho_{\tau}(r) +  \lambda\|D\beta\|_1+g(w)+\frac{\gamma}{2}\|A_{1}\beta+A_{2}r+A_{3}w-c+u\|_{2}^{2},
\end{align*}
where $u=(u_{1}^{T},u_{2}^{T},u_{3}^{T})^{T}$ is the scaled dual variable satisfying $u_{1} \in \mathbb{R}^{n}$,
$u_{2} \in \mathbb{R}^{q}$, $u_{3} \in \mathbb{R}^{s}$.
The updating rules are as follows,
\begin{align*}
\beta^{(k+1)}
&=\arg\min_{\beta}  \lambda\|D\beta\|_1+\frac{\gamma}{2}\| A_{1}\beta+A_{2}
r^{(k)}+A_{3}w^{(k)}-c+u^{(k)}\|_{2}^{2},\\
r_{i}^{(k+1)}
&=[y_{i}-x_{i}^{T}\beta^{(k+1)}-u_{1i}^{(k)}-\tau/\gamma]_{+}-
[-y_{i}+x_{i}^{T}\beta^{(k+1)}+u_{1i}^{(k)}+(\tau-1)/\gamma]_{+},\\
w^{(k+1)}&=(C\beta^{(k+1)}-d+u_{2}^{(k)})_{+},\\
u^{(k+1)}&=u^{(k)}+
\begin{pmatrix}
 X\beta^{(k+1)}+r^{(k+1)}-y \\
 C\beta^{(k+1)}-w^{(k+1)}-d \\
 E\beta^{(k+1)}-f
 \end{pmatrix}.
 \end{align*}
 The update for $\beta$ is a general lasso problem and needs an iterative computation such as
 coordinate descent method. This may cause the inner iteration during the ADMM update steps.

\subsubsection*{ADMM3.Proj}
By introducing new variables $r \in \mathbb{R}^{n}$, $w \in \mathbb{R}^{p}$, we formulate the problem as
\begin{align*}
& \rho_{\tau}(r) + \lambda\|D\beta\|_1 + \phi(w) \nonumber \\
&\text{subject to \
  $y - X\beta = r$, \ $\beta = w$},
\end{align*}
where $\phi(w) = 0$ if $Cw \geq d$ and $Ew = f$ and $= \infty$ otherwise.
Denote \begin{align*}
A_{1}=
 \begin{pmatrix}
 X \\
 I_{p}
 \end{pmatrix}, \quad
A_{2}
 = \begin{pmatrix}
  I_{n} \\
  0_{p}
 \end{pmatrix},\quad
A_{3}
 = \begin{pmatrix}
 0_{n} \\
 -I_{p}
 \end{pmatrix}\quad
c=\begin{pmatrix}
  y\\
  0_{p}
 \end{pmatrix}.
\end{align*}
The augmented Lagrangian
is
\begin{align*}
\mathcal{L}_{\gamma}(\beta,r,w,u)&= \rho_{\tau}(r) +  \lambda\|D\beta\|_1+f(w)+\frac{\gamma}{2}\|A_{1}\beta+A_{2}r+A_{3}w-c+u\|_{2}^{2},
\end{align*}
where $u=(u_{1}^{T},u_{2}^{T})^{T}$ is the scaled dual variable satisfying $u_{1} \in \mathbb{R}^{n}$, $u_{2} \in \mathbb{R}^{p}$.

The updating rules are as follows,
\begin{align*}
\beta^{(k+1)}
&=\arg\min_{\beta}  \lambda\|D\beta\|_1+\frac{\gamma}{2}\| \begin{pmatrix}
 X \\
 I_{p}
 \end{pmatrix}\beta+\begin{pmatrix}
  I_{n} \\
  0_{p}
 \end{pmatrix}r^{(k)}+\begin{pmatrix}
 0_{n} \\
  -I_{p}
 \end{pmatrix}w^{(k)}-\begin{pmatrix}
  y\\
  0_{p}
 \end{pmatrix}+u^{(k)}\|_{2}^{2},\\
r_{i}^{(k+1)}
&=[y_{i}-x_{i}^{T}\beta^{(k+1)}-u_{1i}^{(k)}-\tau/\gamma]_{+}-
[-y_{i}+x_{i}^{T}\beta^{(k+1)}+u_{1i}^{(k)}+(\tau-1)/\gamma]_{+},\\
w^{(k+1)}&=\text{Proj$_{\mathcal{C}}$}(\beta^{(k+1)}+u_{2}^{(k)}),\\
u^{(k+1)}&=u^{(k)}+
\begin{pmatrix}
 X\beta^{(k+1)}+r^{(k+1)}-y \\
 \beta^{(k+1)}-w^{(k+1)}
 \end{pmatrix}.
\end{align*}
The update for $w$ is a projection on set $\mathcal{C}=\{w: Cw\geq d, Ew=f \}$. The update for $\beta$ is a general lasso problem and needs an iterative computation such as
 coordinate descent method.

\section*{ Appendix C}
\subsection*{ The update of $z$ for MCP and SCAD penalty}
We derive in details how to update variable $z$ for the ADMM4.Constr algorithm with the SCAD and MCP penalty. SCAD and MCP penalties are asymptotically unbiased and are more aggressive in enforcing a sparser solution. The MCP has the form
$$p_{\lambda,\xi}(t)=\lambda \int_{0}^{t}\min\{1,(\xi-x/\lambda)_{+}/(\xi-1)\}dx,\ \xi>2,$$
and the SCAD penalty is
$$p_{\lambda,\xi}(t)=\lambda \int_{0}^{t}(1-x/(\xi \lambda))_{+}dx, \ \xi>1,$$
where $\xi$ is a parameter that controls the concavity of the penalty function.

The $z$-update can be done for each element $z_{j}$ for $j=1,\ldots,m$ with SCAD and MCP penalty
\begin{align*}
z_{j}^{(k+1)}=\arg\min_{z_{j}}\frac{\gamma}{2}( \delta_{j}^{(k)}-z_{j})^{2}+p_{\lambda,\xi}(z_{j}),
\end{align*}
 where $\delta_{j}^{(k)}=D_{j}\beta^{(k+1)}+u_{2j}^{(k)}$.
 For the SCAD penalty, the objective function for updating $z_{j}$ is convex when $\xi >1/\gamma+1$. Hence, the closed-form solution for SCAD penalty with $\xi>1/\gamma+1$ is
\begin{equation*}
z_{j}^{(k+1)}= \left\{
             \begin{array}{cl}
             \text{ST}(\delta_{j}^{(k)},\lambda/\gamma) &    {\mbox{if}\  |\delta_{j}^{(k)}|\leq \lambda+\lambda/\gamma}, \\
             \frac{\text{ST}(\delta_{j}^{(k)},\xi\lambda/((\xi-1)\gamma))}
             {1-1/((\xi-1)\gamma)}&    {\mbox{if}\  \lambda+\lambda/\gamma <|\delta_{j}^{(k)}|\leq \xi\lambda}, \\
             \delta_{j}^{(k)}  &{\mbox{if}\  |\delta_{j}^{(k)}|>\xi\lambda},
             \end{array}
        \right.
\end{equation*}
where $\text{ST}(t, \lambda)=\text{sign}(t)(|t|-\lambda)_{+}$ is the soft thresholding rule,
 and $(x)_{+}=x$ if $x>0$, and $(x)_{+}=0$ otherwise.

For the MCP penalty, the objective function for updating $z_{j}$ is convex when $\xi >1/\gamma$. Hence, the closed-form solution for the MCP penalty with $\xi >1/\gamma$ is
\begin{equation*}
z_{j}^{(k+1)}= \left\{
             \begin{array}{cl}
             \frac{\text{ST}(\delta_{j}^{(k)},\lambda/\gamma)}{1-1/\xi\gamma}&    {\mbox{if}\  |\delta_{j}^{(k)}|\leq \xi\lambda}, \\
             \delta_{j}^{(k)}  &{\mbox{if}\  |\delta_{j}^{(k)}|> \xi\lambda}.
             \end{array}
        \right.
\end{equation*}

\subsection*{The stopping criterion of parallel ADMM algorithm}
According to Boyd et al.(2010), the primal and dual residuals for the $m$th subset are
  \begin{align*}
  r_{\text{pri},m}^{k+1}
   =\begin{pmatrix}
 X_{m}\beta_{m}^{(k+1)}+r_{m}^{(k+1)}-y_{m} \\
 D\beta_{m}^{(k+1)}-z^{(k+1)} \\
 C\beta_{m}^{(k+1)}-w^{(k+1)}-d \\
 E\beta_{m}^{(k+1)}-f \\
 \beta_{m}^{(k+1)}-\beta^{(k+1)}
 \end{pmatrix},
   \end{align*}
 \begin{align*}
s_{m}^{k+1}&=\gamma \big(X_{m}^{T}(r_{m}^{(k+1)}-r_{m}^{(k)})- D^{T}(z^{(k+1)}-z^{(k)})-C^{T}(w^{(k+1)}-w^{(k)})- (\beta^{(k+1)}-\beta^{(k)})\big).
\end{align*}
The primal and dual residuals for the parallel ADMM is the aggregation of all such terms,
 \begin{align*}
r_{\text{pri}}^{(k+1)}=\sqrt{\sum_{m=1}^{M}\|r_{\text{pri},m}^{k+1}\|_{2}^{2}},\quad
s^{(k+1)}= \sqrt{\sum_{m=1}^{M}\|s_{m}^{k+1}\|_{2}^{2}}.
  \end{align*}
  The stopping rule is set as
\begin{align*}
r_{\text{pri}}^{(k+1)}\leq \epsilon^{\text{pri}}, \quad s^{(k+1)}\leq \epsilon^{\text{dual}}.
\end{align*}
where
 \begin{align*}
 \epsilon^{\text{prim}} &=\sqrt{n+M(m+q+s+p)}\epsilon ^{\text{abs}}
 +\epsilon ^{\text{rel}}\max\big\{\sqrt{\sum_{m=1}^{M}\|A_{1,m}\beta_{m}^{(k+1)}\|^{2}_{2}}, \sqrt{\sum_{m=1}^{M}\|r_{m}^{(k+1)}\|^{2}_{2}}, \\
 &\sqrt{M}\|z^{(k+1)}\|_{2},
\sqrt{M}\|w^{(k+1)}\|_{2},
 \sqrt{M}\|\beta^{k+1}\|_{2},
 \sqrt{\sum_{m=1}^{M}\|b_{m}^{(k+1)}\|^{2}_{2}}\big\},\\
 \epsilon^{\text{dual}} &=\sqrt{Mp}\epsilon ^{\text{abs}}+\epsilon ^{\text{rel}}\|\sqrt{\sum_{m=1}^{M}\|\gamma A_{1,m}u_{m}^{(k+1)}\|^{2}_{2}},
 \end{align*}
and
 \begin{align*}
A_{1,m}=(X_{m}^{T},D^{T},C^{T},E^{T},I_{p}^{T})^{T},\quad
 b_{m}=(y^{T},0,d^{T},f^{T},0)^{T}.
\end{align*}


\end{document}